\pdfoutput=1
\documentclass[11pt]{article}
\usepackage[table]{xcolor} 
\usepackage{tikz}
\usepackage{pgfplots}
\pgfplotsset{compat=1.18}
\usetikzlibrary{shapes.geometric, arrows}
\usepackage{amsmath}
\usepackage{amssymb}
\usepackage{amsfonts}
\usepackage{hyperref}
\usepackage{algorithm}
\usepackage[noend]{algpseudocode}

\tikzstyle{block} = [rectangle, rounded corners, minimum width=3cm, minimum height=1cm, text centered, draw=black, fill=blue!30]
\tikzstyle{arrow} = [thick,->,>=stealth]
\tikzstyle{input} = [ellipse, draw, fill=green!30, minimum height=1cm, minimum width=1.5cm, text centered]

\title{On Bitcoin Price Prediction}
\usepackage[utf8]{inputenc} % encode en UTF8

\usepackage[english]{babel} % langue francaise

\usepackage[T1]{fontenc}

\usepackage{caption}

\usepackage{graphicx}

\usepackage[top=2cm, bottom=2cm, left=2cm, right=2cm]{geometry}

\usepackage{amsmath,amsfonts,amssymb}

\usepackage{array}

\usepackage{fancyhdr}

\usepackage{lastpage}

\usepackage[toc,page]{appendix}

\usepackage{algorithm}
\usepackage[noend]{algpseudocode}

\usepackage{ragged2e}
\usepackage{amssymb, bm}
\usepackage{float}
\usepackage{graphicx}

\usepackage{listings}

\usepackage{siunitx}
\usepackage{calrsfs}
\usepackage{titlesec}
\DeclareMathAlphabet{\pazocal}{OMS}{zplm}{m}{n}

\DeclareFixedFont{\ttb}{T1}{txtt}{bx}{n}{12} % for bold
\DeclareFixedFont{\ttm}{T1}{txtt}{m}{n}{12}  % for normal

\definecolor{deepblue}{rgb}{0,0,0.5}
\definecolor{deepred}{rgb}{0.6,0,0}
\definecolor{deepgreen}{rgb}{0,0.5,0}
\definecolor{comment}{rgb}{0.55,0.11,0.46}
\usepackage[utf8]{inputenc}
\usepackage{listings}

\definecolor{codegreen}{rgb}{0,0.6,0}
\definecolor{codegray}{rgb}{0.5,0.5,0.5}
\definecolor{codepurple}{rgb}{0.960, 0.356, 0}
\definecolor{backcolour}{rgb}{0.97, 0.97, 0.97}

\lstdefinestyle{mystyle}{
  backgroundcolor=\color{backcolour},   commentstyle=\color{codegreen},
  keywordstyle=\color{blue},
  numberstyle=\tiny\color{codegray},
  stringstyle=\color{codepurple},
  basicstyle=\ttfamily\tiny,
  breakatwhitespace=false,         
  breaklines=true,                 
  captionpos=b,                    
  keepspaces=true,                 
  numbers=left,                    
  numbersep=5pt,                  
  showspaces=false,                
  showstringspaces=false,
  showtabs=false,                  
  tabsize=2
}
\newcommand\pythonstyle{\lstset{
language=Python,
basicstyle=\ttm,
morekeywords={self},              % Add keywords here
keywordstyle=\ttb\color{deepblue},
emph={MyClass,__init__},          % Custom highlighting
emphstyle=\ttb\color{deepred},    % Custom highlighting style
stringstyle=\color{deepgreen},
frame=tb,                         % Any extra options here
showstringspaces=false,
commentstyle = \ttm\color{comment}
}}

\lstnewenvironment{python}[1][]
{
\pythonstyle
\lstset{#1}
}
{}

\hypersetup{
  colorlinks   = true, %Colours links instead of ugly boxes
  urlcolor     = blue, %Colour for external hyperlinks
  linkcolor    = black, %Colour of internal links
  citecolor   = red %Colour of citations
}

\begin{document}
\renewcommand{\tablename}{Table}
\newcommand\numberthis{\addtocounter{equation}{1}\tag{\theequation}}
\newcommand{\HRule}{\rule{\linewidth}{1mm}}

%%%%%%%%%%%%%%%%%%%%%%%%%%%%%%%%%%%%%%%%%%%%%%%%%%%%%
\def\TP{On Bitcoin Price Prediction}
\def\Redac{Grégory \textsc{Bournassenko\footnote{gregory.bournassenko@etu.u-paris.fr}}}
\def\Date{September 12, 2024}
\def\Encadrant{Mariana \textsc{Rojas Breu}}
% tp.tex — Page de garde style article scientifique

\pagestyle{empty}

\begin{center}
\vspace*{1cm}

{\Large \textbf{\TP}} \\[2em]

{\normalsize \textbf{\Redac}} \\[0.5em]
Université Paris Cité \\[1.5em]

% \textbf{Supervisor:} \Encadrant \\[0.3em]
% \Date \\[2.5em]
\end{center}

\vspace{2em}

\begin{center}
\begin{minipage}{0.85\textwidth}
\justifying
In recent years, cryptocurrencies have attracted growing attention from both private investors and institutions. Among them, Bitcoin stands out for its impressive volatility and widespread influence. This paper explores the predictability of Bitcoin’s price movements, drawing a parallel with traditional financial markets. We examine whether the cryptocurrency market operates under the efficient market hypothesis (EMH) or if inefficiencies still allow opportunities for arbitrage. Our methodology combines theoretical reviews, empirical analyses, machine learning approaches, and time series modeling to assess the extent to which Bitcoin’s price can be predicted. We find that while, in general, the Bitcoin market tends toward efficiency, specific conditions, including information asymmetries and behavioral anomalies, occasionally create exploitable inefficiencies. However, these opportunities remain difficult to systematically identify and leverage. Our findings have implications for both investors and policymakers, particularly regarding the regulation of cryptocurrency brokers and derivatives markets.
\end{minipage}
\end{center}

\vfill

%%%%%%%%%%%%%%%%%%%%%%%%%%%%%%%%%%%%%%%%%%%%%%%%%%%%%

%\newpage
%\strut
\newpage
\cleardoublepage
\pagenumbering{roman} 
%\tableofcontents \newpage \strut \cleardoublepage

%%%%%%%%%%%%%%%%%%%%%%%%%%%%%%%%%%%%%%%%%%%%%%%%%%%%%

\pagenumbering{arabic}

\pagestyle{fancy}
\renewcommand{\headrulewidth}{0pt}
\renewcommand\footrulewidth{0.1pt}
\fancyhead[R]{}
\fancyhead[L]{}
\fancyfoot[R]{\thepage/\pageref{LastPage}}
\fancyfoot[L]{\TP}
\fancyfoot[C]{ }

%%%%%%%%%%%%%%%%%%%%%%%%%%%%%%%%%%%%%%%%%%%%%%%%%%%%%
\setcounter{page}{1}

\tableofcontents
\cleardoublepage
\listoffigures
\listoftables
\cleardoublepage

\section{Introduction}

The price of Bitcoin has lost almost 50\% of its value since last November, almost as much as Orpea's stock value after its scandal. In Orpea's case, the correlation is clear with the scandal, but for Bitcoin, such irrational volatility is rather usual.

\begin{figure}[H]
\centering
\begin{minipage}{.5\textwidth}
  \centering
  \includegraphics[width=1\linewidth]{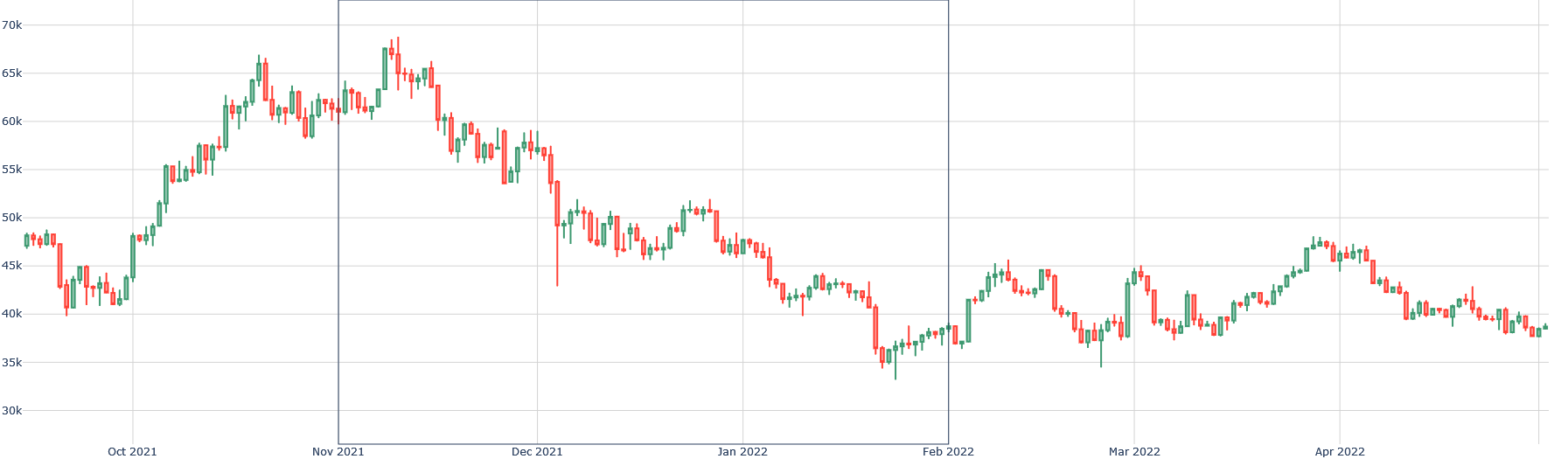}
  \caption{$\blacktriangledown 50\%$ BTC/USD [11/2021-02/2022]}
  \label{fig:1}
\end{minipage}%
\begin{minipage}{.5\textwidth}
  \centering
  \includegraphics[width=1\linewidth]{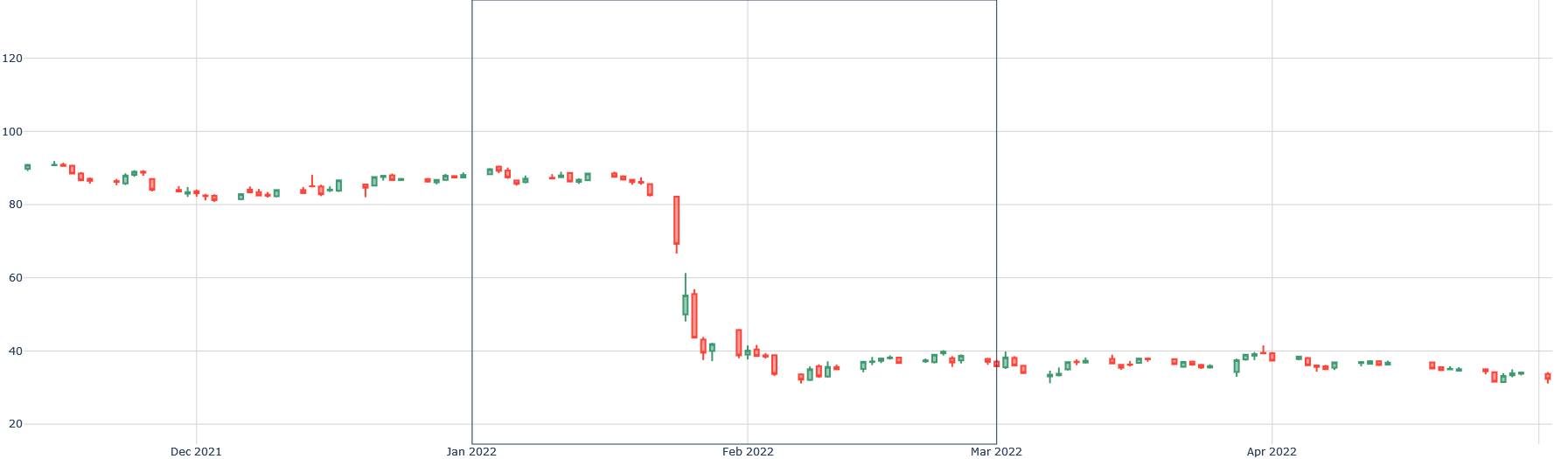}
  \caption{$\blacktriangledown 50\%$ ORP [01/2022-03/2022]}
  \label{fig:2}
\end{minipage}
\end{figure}

The notion of prediction is vague, especially regarding price prediction: isn't price itself the result of agents' predictions about the value of an asset? Are we therefore predicting a prediction? For simplicity, we will use the term prediction as defined by American economist Alfred Cowles in his paper \cite{cowles1933can}, particularly in the second part, where he analyzes the reliability of "forecasters" on stock market volatility. Bitcoin, for its part, is a decentralized cryptocurrency, created in 2008, based on a "proof of work" mining protocol and a robust transaction system as explained by Satoshi Nakamoto \cite{nakamoto2008bitcoin}.

\begin{figure}[H]
    \centering
   \includegraphics[width=1\linewidth]{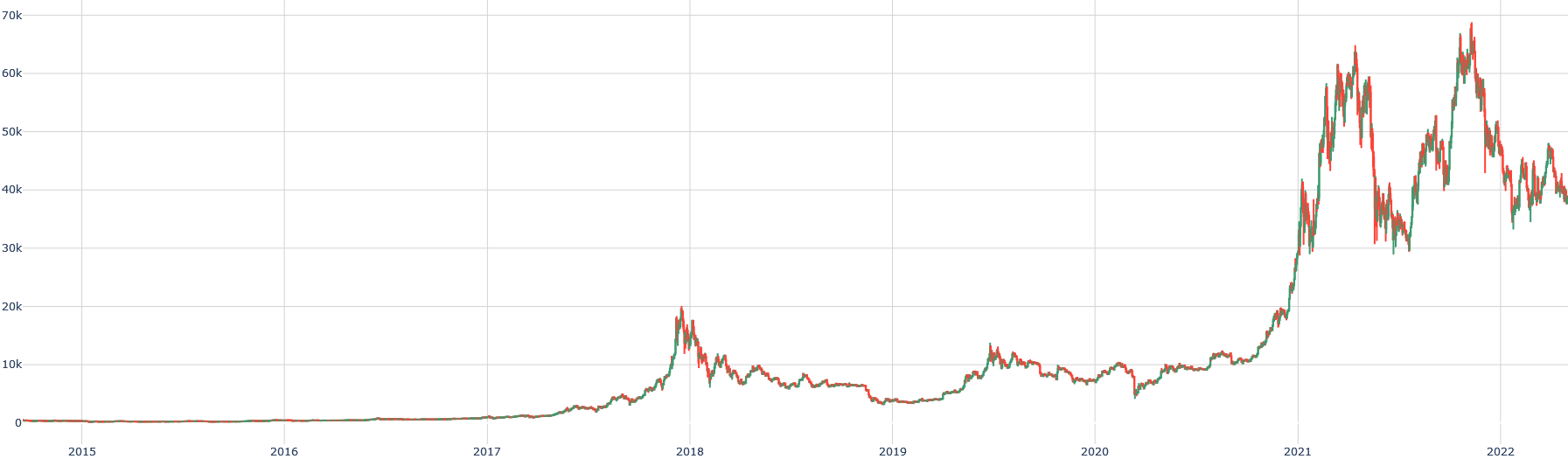}
    \caption{$\blacktriangle 9,000\%$ BTC/USD [2014-2022]}
    \label{fig:3}
\end{figure}

As shown above, Bitcoin has progressively gained success: initially used for anonymous transactions on illegal markets, it became a speculative tool for individuals, and eventually attracted institutional interest, despite limited daily usage \cite{baur2015cryptocurrencies}. Notably, Bitcoin's underlying technology, Blockchain, was actually invented by researchers Haber and Stornetta \cite{haber1990time}, not Nakamoto, although Nakamoto was the first to apply it at large scale.

The literature on cryptocurrency prediction remains relatively poor, given the recent emergence of the technology. Virtually no academic papers referenced cryptocurrencies before 2008. Instead, much research focuses on machine learning techniques for cryptocurrency prediction. However, similarities with financial markets exist (closer to forex than stocks due to the monetary nature of cryptocurrencies), a domain extensively studied since the early 1900s. From Louis Bachelier's Gaussian model \cite{bachelier1900theorie} to Mathieu Rosenbaum's rough Heston model \cite{gatheral2018volatility}, and Gordon-Shapiro's valuation model \cite{gordon1956capital}, numerous theories have been proposed. Yet, debates persist regarding market behavior.

According to Eugene Fama \cite{fama1970efficient}, a rational market cannot be systematically beaten. Louis Bachelier \cite{bachelier1900theorie} states, "The determination of these activities depends on an infinite number of factors: therefore, a precise mathematical forecast is absolutely impossible." Nevertheless, Keynes \cite{keynes1937general} compared the market to a beauty contest: predicting what the majority will find beautiful, not objective beauty itself. This idea echoes momentum strategies and aligns with Charles Dow's technical analysis \cite{brown1998dow}.

Alternatively, Warren Buffett promotes stock-picking and value investing, diverging from Markowitz's modern portfolio theory \cite{steinbach2001markowitz}. However, Buffett's method, focusing on selecting promising assets, differs from our study, where the asset (Bitcoin) is preselected. Burton Malkiel \cite{malkiel2003efficient} famously claimed that "a blindfolded monkey throwing darts at a newspaper's financial pages could perform as well as professional investors," although empirical studies \cite{pernagallo2020blindfolded} challenge this assertion.

To explore random versus selected portfolios, we define a Python function \texttt{isRandomBetter($\Omega, n, k$)} (code in Appendix \ref{appendix:isRandomBetter}). Results:

\begin{table}[H]
\centering
\begin{tabular}{ccccccc}
\hline
Test No. & $|\Omega|$ & $\overline{R_{\Omega}}$ & $n$ & $k$ & \% $\omega$ better than $\Omega$ & Result\\ \hline
1 & 141 & 998 & 10 & 10 & 20\% & \texttt{False}\\ \hline
2 & 141 & 998 & 10 & 20 & 30\% & \texttt{False}\\ \hline
3 & 141 & 998 & 20 & 10 & 40\% & \texttt{False}\\ \hline
4 & 141 & 998 & 20 & 20 & 30\% & \texttt{False}\\ \hline
5 & 141 & 998 & 20 & 30 & 60\% & \texttt{True}\\ \hline
6 & 141 & 998 & 30 & 20 & 57\% & \texttt{True}\\ \hline
7 & 141 & 998 & 30 & 30 & 47\% & \texttt{False}\\ \hline
8 & 141 & 998 & 30 & 10 & 27\% & \texttt{False}\\ \hline
9 & 141 & 998 & 10 & 30 & 20\% & \texttt{False}\\ \hline
10 & 141 & 998 & 40 & 5 & 25\% & \texttt{False}\\ \hline
\end{tabular}
\caption{Results of \texttt{isRandomBetter($\Omega, n, k$)}}
\label{tab:isRandomBetter}
\end{table}

Choosing a random crypto portfolio in 2021 was not optimal.

We will investigate whether Bitcoin price predictability depends on market efficiency. Given the cryptocurrency market's heterogeneity, various scenarios (competitive markets, manipulated markets, rational/irrational agents) are expected.

We will show that, by default, the crypto market tends to be efficient, although inefficiencies sometimes appear, albeit difficult to exploit systematically.

We will address prediction methods under efficient market conditions, focusing on time series analysis and machine learning algorithms. We will also study prediction under inefficiency contexts, emphasizing empirical observations and stylized facts.

Let's first examine the case when the market is efficient.

\section{The Cryptocurrency Market is Efficient}

We first assume an efficient market. We will explain the concept's origins, assumptions, verify some of them, discuss model evolutions, and their implications for cryptocurrencies. We will also analyze this through machine learning and quantitative techniques, reflecting critically on the results.

\subsection{Eugene Fama and the Notion of No Arbitrage Opportunities}

We start with Fama's \cite{fama1970efficient} definition of efficient markets, comparing the US stock market and cryptocurrencies. Fama's idea implies no arbitrage opportunities. However, as we will see later, arbitrage is relatively common in crypto markets (price differences between brokers).

\begin{figure}[H]
\centering
\begin{minipage}{.5\textwidth}
  \centering
  \includegraphics[width=1\linewidth]{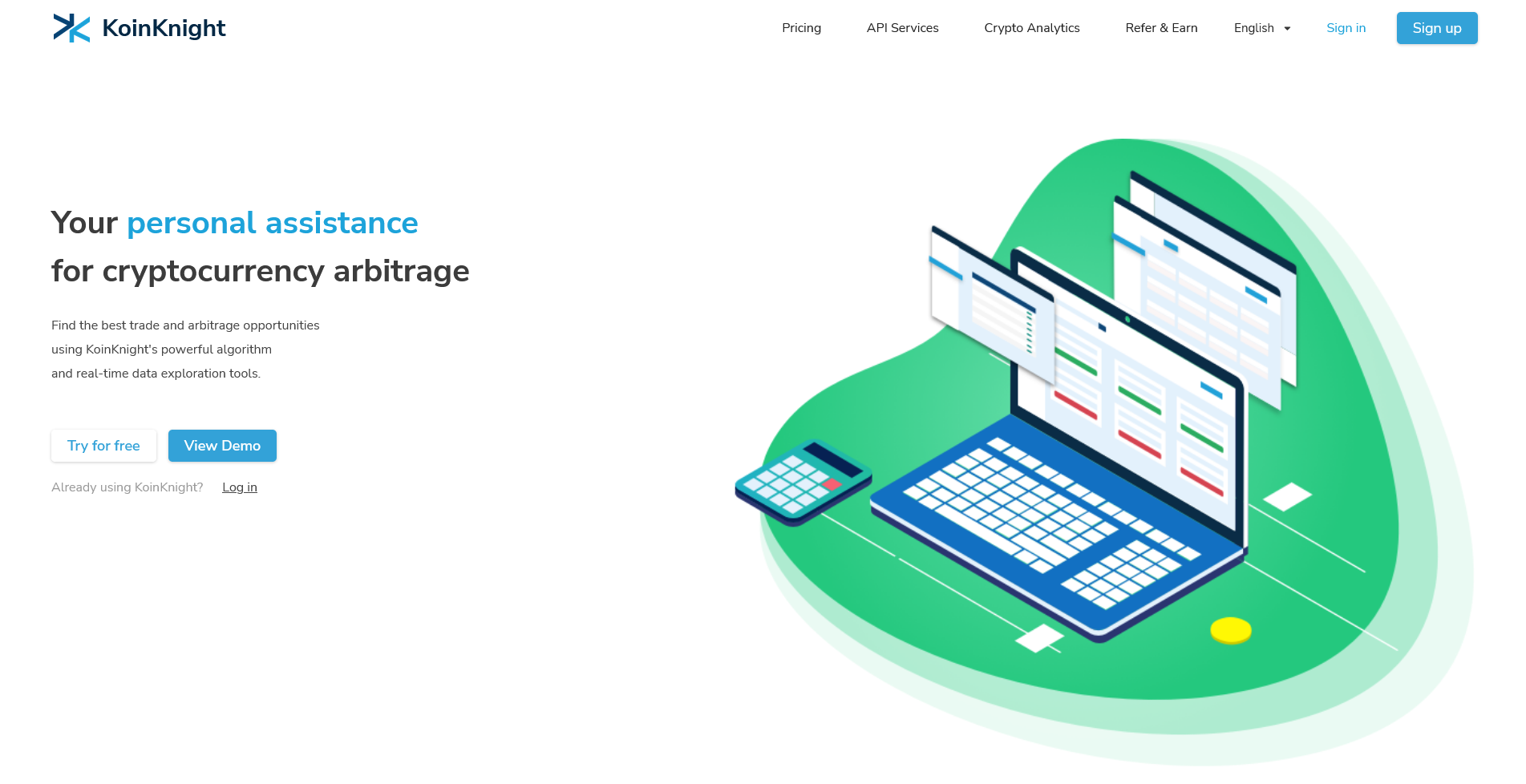}
  \caption{KoinKnight}
  \label{fig:4}
\end{minipage}%
\begin{minipage}{.5\textwidth}
  \centering
  \includegraphics[width=1\linewidth]{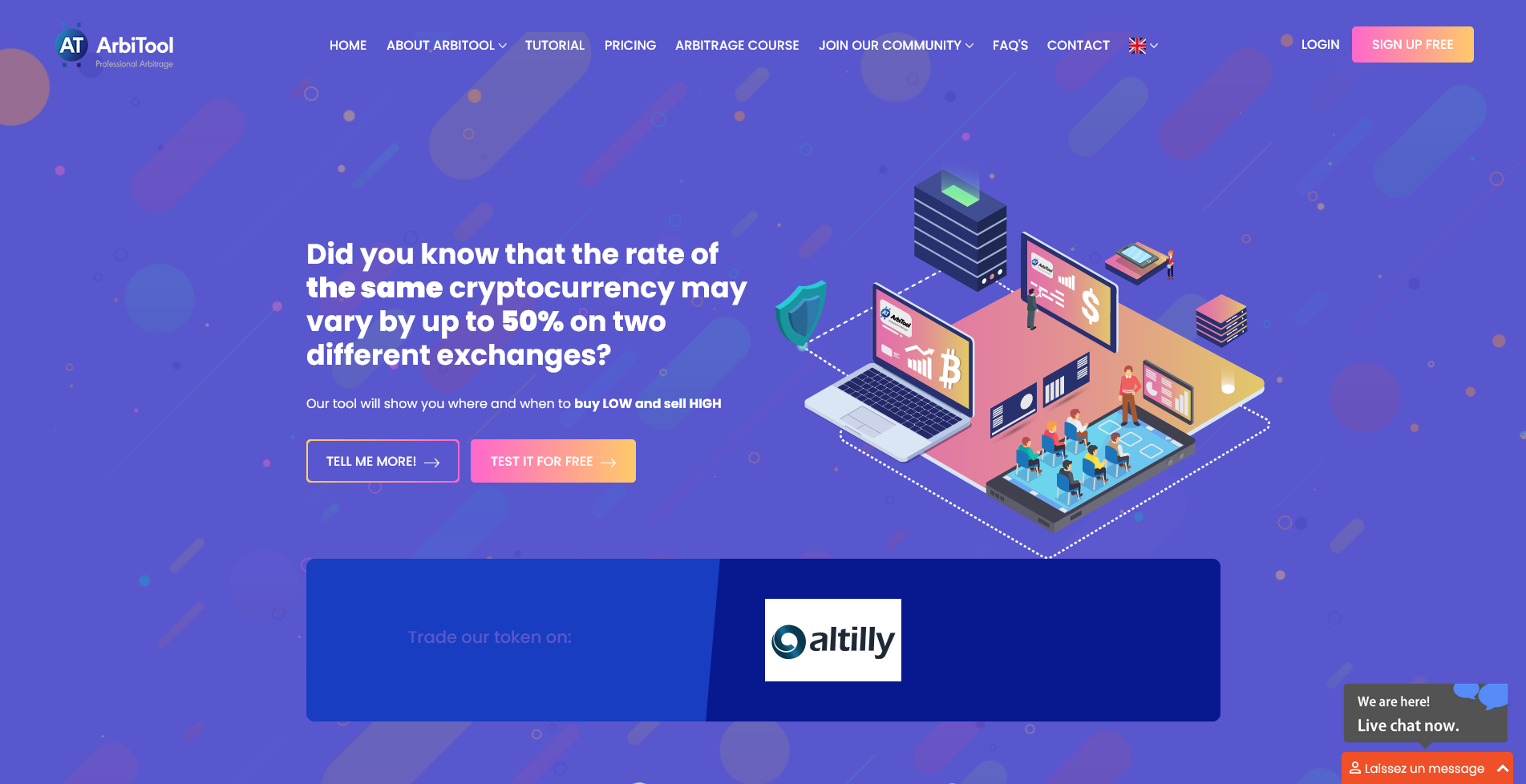}
  \caption{ArbiTool}
  \label{fig:5}
\end{minipage}
\end{figure}

At a discretionary level, however, arbitrage opportunities are rarely exploitable due to transfer fees and liquidity issues.

\subsubsection{Efficient Market Hypothesis Adaptation to Cryptocurrencies}

Fama \cite{fama1970efficient} outlined several conditions for market efficiency and its three forms. Let's check them for crypto markets.

First, agents should be rational. In crypto, this is unlikely. For example, Dogecoin rose by 14,000\% mainly due to memes and social media \cite{chohan2021history}:

\begin{figure}[H]
    \centering
   \includegraphics[width=1\linewidth]{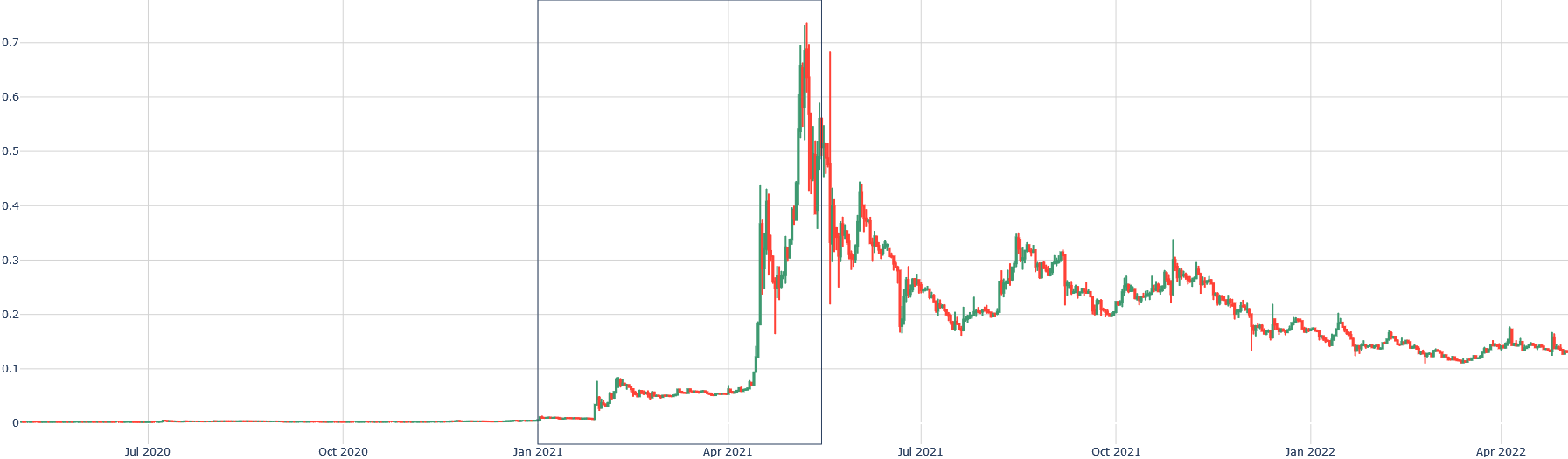}
    \caption{$\blacktriangle 14,000\%$ DOGE/USD [01/2021-05/2021]}
    \label{fig:6}
\end{figure}

Individuals should not influence the market. Elon Musk, however, can shift prices with a single tweet:

\begin{figure}[H]
\centering
\begin{minipage}{.5\textwidth}
  \centering
  \includegraphics[width=0.3\linewidth]{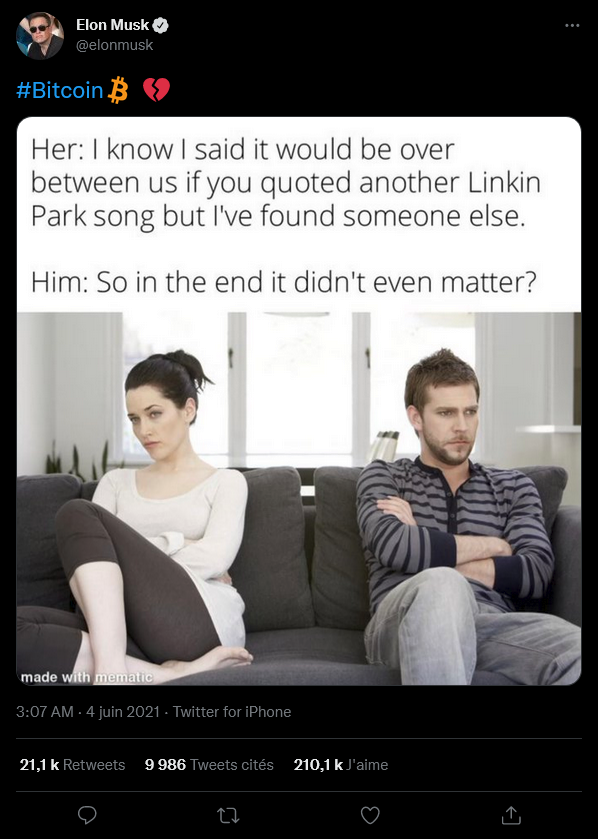}
  \caption{Negative tweet on 04/06/2021}
  \label{fig:7}
\end{minipage}%
\begin{minipage}{.5\textwidth}
  \centering
  \includegraphics[width=1\linewidth]{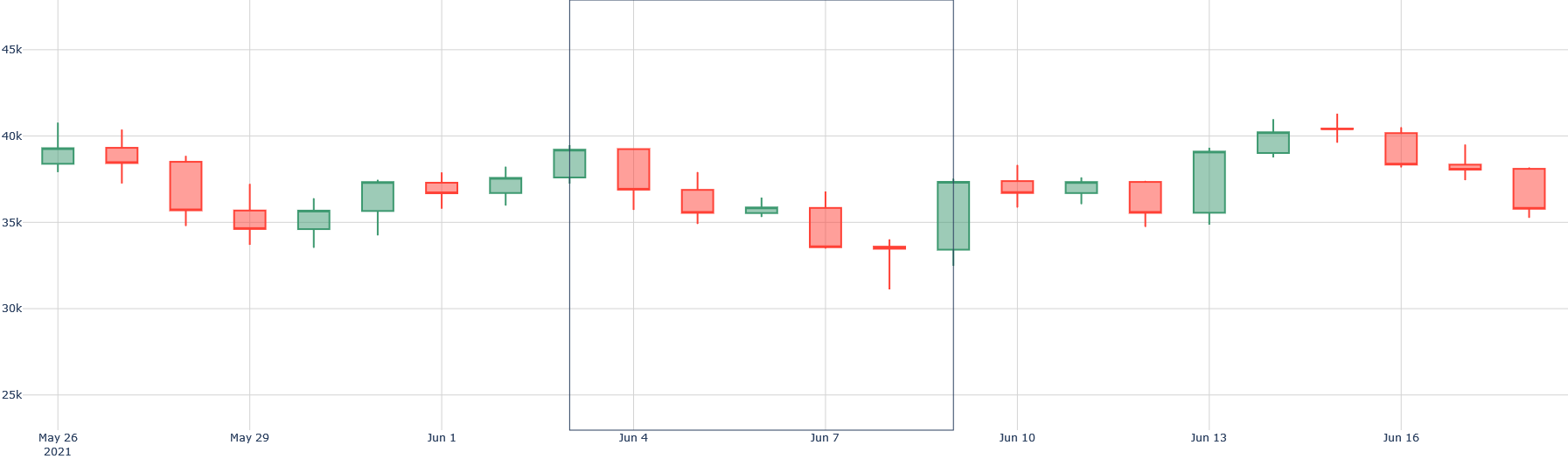}
  \caption{Observed correlation: $\blacktriangledown 15\%$ BTC/USD [04/06/2021-08/06/2021]}
  \label{fig:8}
\end{minipage}
\end{figure}

No information asymmetry should exist. Yet, insider knowledge (e.g., hacks) creates advantages \cite{biais2020equilibrium}.

Information should be free. For crypto, public data is widely available, though high-frequency trading data is costly \cite{grossman1976information}.

Taxes should be low. Given international diversity, this varies.

Regarding efficiency forms:

\textbf{Strong form}: all public and private info is priced. However, events like Binance's launch in 2017 or the Bitconnect scandal in 2018 show that insiders could have benefited:

\begin{figure}[H]
    \centering
   \includegraphics[width=1\linewidth]{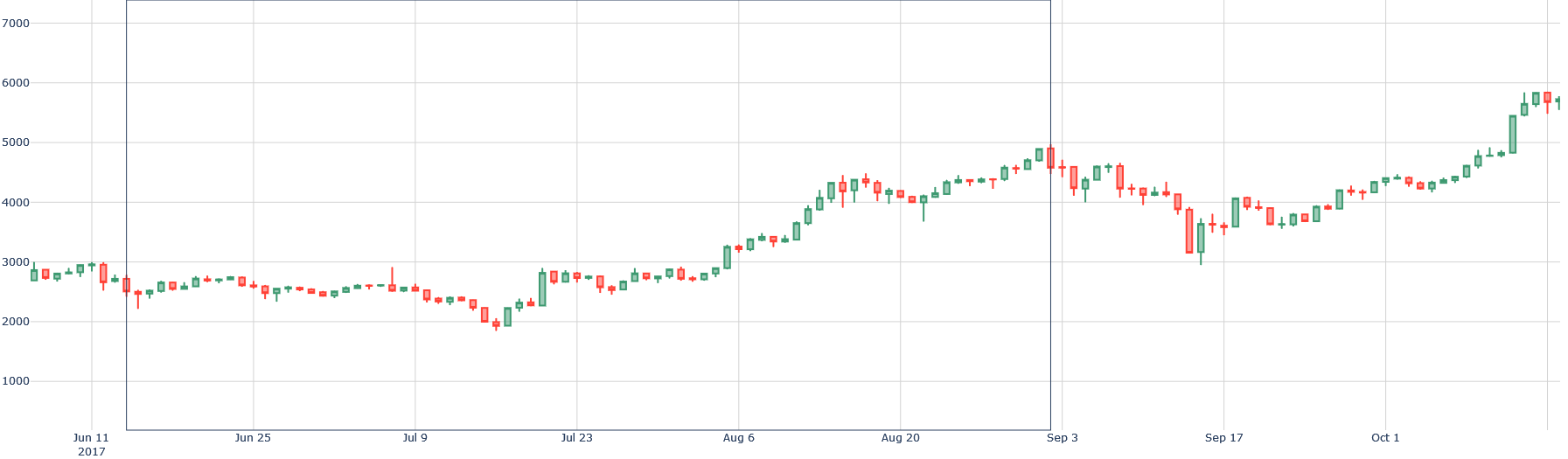}
    \caption{$\blacktriangle 150\%$ BTC/USD [13/06/2017-01/09/2017]}
    \label{fig:9}
\end{figure}

\begin{figure}[H]
    \centering
   \includegraphics[width=1\linewidth]{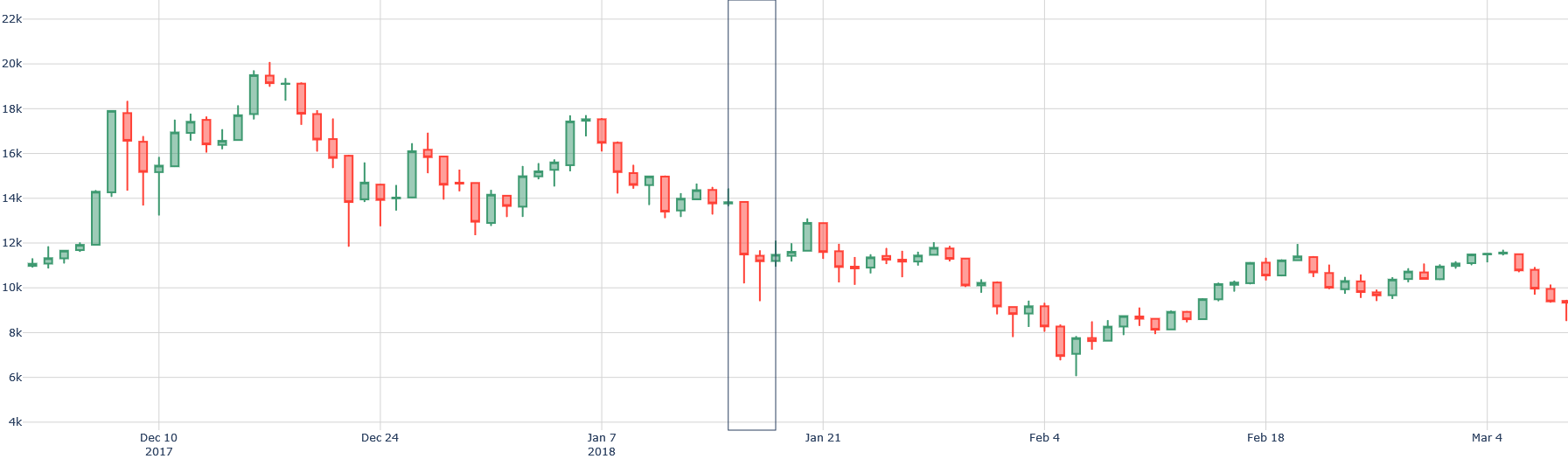}
    \caption{$\blacktriangledown 15\%$ BTC/USD [16/01/2018-17/01/2018]}
    \label{fig:10}
\end{figure}

\textbf{Semi-strong form}: all public info is priced. The crypto market reacts quickly to news, as seen with Coinbase's NASDAQ listing:

\begin{figure}[H]
    \centering
   \includegraphics[width=1\linewidth]{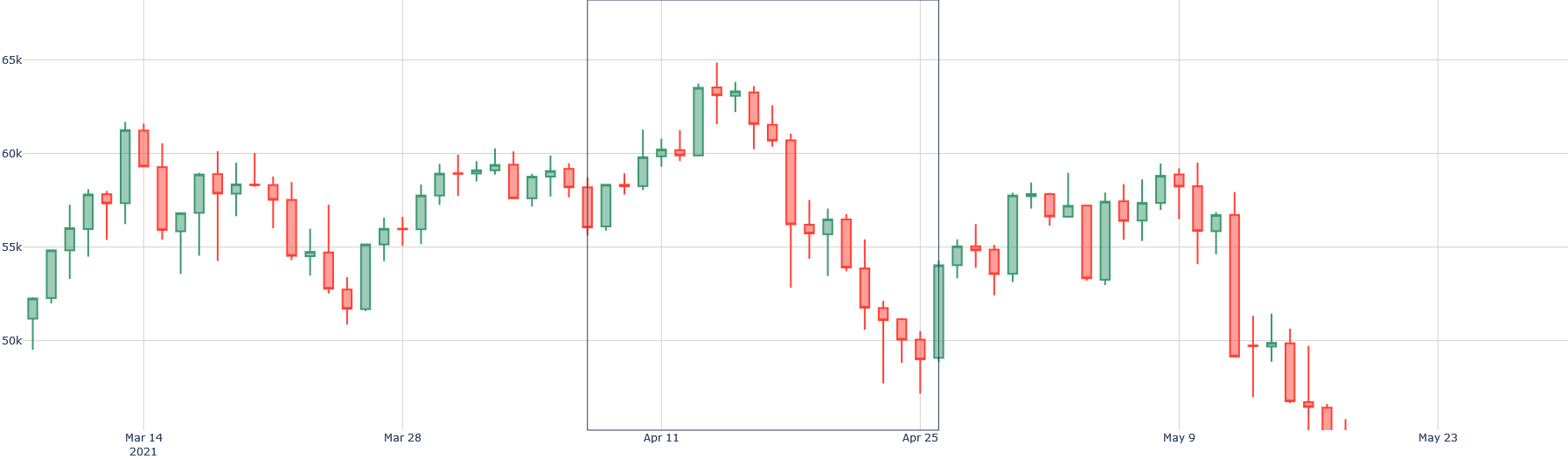}
    \caption{$\blacktriangledown 22\%$ BTC/USD [14/04/2021-25/04/2021]}
    \label{fig:11}
\end{figure}

The day before its IPO, BTC/USD increased by almost 7\%, before losing more than 20\% ten days later. The weak form assumes that all historical price information is already reflected in the current price. This form challenges technical analysis, which specializes precisely in analyzing past returns. These analyses are widely shared on social media, due to their ease of implementation, and attract a (too?) proselytizing community. The idea is to use indicators mainly based on past fluctuations to make future predictions. Among the usual indicators (according to the \href{https://ta-lib.org/}{TA-Lib} library, considered a reference) are: RSI (Relative Strength Index), SMA (Simple Moving Average), BBANDS (Bollinger Bands). Let us check, for example, whether a "mean-reversion" strategy would be more effective than a simple "hold" (buy-sell only once) and more effective than a random strategy by backtesting these strategies on 2021. If not, we could conjecture that, over the entire year of 2021, it was useless to use a "mean-reversion" strategy (which assumes that when the current price is too "far" from the moving average (SMA), the price will return to its "mean")). This may also give us an indication about the market efficiency form.

We will base our analysis on a set $\Omega$ of crypto-assets. For each element in $\Omega$, we will test three strategies: mean-reversion, hold, and random. We assume short-selling is allowed. Let $P_t$ be the price at time $t$, $M_t(n)$ the moving average at time $t$ with a window of $n$ days, $\omega_i$ the $i^{th}$ element of $\Omega$, and $r \in [0,100]$ a percentage around $M_t(n)$ indicating the threshold at which we open/close a position. The mean-reversion strategy will be constructed as follows: if $P_t > M_t(n)+(\frac{M_t(n)\times r }{100})$, then sell $\omega_i$ at price $P_t$; if $P_t < M_t(n)-(\frac{M_t(n)\times r }{100})$, then buy $\omega_i$ at price $P_t$, with $t$ ranging from [01/01/2021, 31/12/2021].

The hold strategy will be constructed as follows: if $t = 01/01/2021$, then buy $\omega_i$ at price $P_t$; if $t = 31/12/2021$, then sell $\omega_i$ at price $P_t$.

The random strategy will be constructed as follows: generate a signal $S \in [\text{buy, sell, hold}]$ with $P(S = \text{buy}) = P(S = \text{sell}) = P(S = \text{hold}) = \frac{1}{3}$. For each $\omega_i$ and for each $t$, if $S=\text{"buy"}$ we buy $\omega_i$ at price $P_t$, if $S=\text{"sell"}$ we sell $\omega_i$ at price $P_t$, if $S=\text{"hold"}$ we do nothing.

Thus, we create a Python function \texttt{isSMABetter($\Omega, n, r$)} that takes as parameters $\Omega$ (the set of crypto-assets), $r$ (the percentage for the SMA thresholds), and $n$ (the window size in days for the SMA), and returns \texttt{True} if the average SMA returns of $\omega_i$ are greater than the average returns of the hold strategy and (strictly) the random strategy in at least 50\% of the cases, and \texttt{False} otherwise.

We only consider daily returns. Indeed, how could we backtest a strategy that only opens positions? We thus place ourselves in a short-term trading scale for each trade, which is consistent with the chartist approach (otherwise, we would prefer a passive investment strategy that requires almost no analysis).

The results of \texttt{isSMABetter($\Omega, n, r$)}, whose code is in Appendix \ref{appendix:isSMABetter}, are as follows:

\begin{table}[H]
\centering
\begin{tabular}{cccccccc}
\hline
$\lvert \Omega \rvert$ & $\overline{R_{\Omega}}$ & $\overline{R_{SMA}}$ & $\overline{R_{Random}}$ & $n$  & $k$  & $(\frac{ \lvert \omega_i \rvert }{\lvert \Omega \rvert})\times 100 / \overline{R_{SMA}} > \overline{R_{\Omega}},\overline{R_{SMA}} > \overline{R_{Random}}$  & Response \\ \hline
116      & 1179   & -484 & -4      & 50 & 20 & 0.00    & \texttt{False}   \\ \hline
\end{tabular}
\caption{Results of \texttt{isSMABetter($\Omega, n, r$)}}
\label{tab:isSMABetter}
\end{table}

It appears that in 2021, among the 116 crypto-assets tested, it was more optimal to have a passive strategy or, at worst, a random strategy, rather than using the moving average in an attempt to generate profits with a day-trading approach (speculation aiming to make a profit within the same day of a market order execution), since the average return obtained with the SMA strategy was the lowest among the three (-484\%), and strictly no crypto-asset (0\%) showed any interest in being traded with an SMA strategy.

We can conjecture that the cryptocurrency market efficiency form is at least weak, and possibly semi-strong, depending on the crypto-assets and periods, but hardly strong.

\subsubsection{Random Walk and Martingale}

In almost all the literature (\cite{lardic2006efficience}, \cite{jovanovic2009modele}...), a random walk is modeled by two elements: the previous observation and white noise. The literature explains that a price can be modeled as: $P_{t+1} = P_t + \varepsilon_{t+1}$, with $\varepsilon = \{\varepsilon_t, t \in N\}$ being white noise. This implies that the best (and only) way to predict the price of an asset is by using its current price. 

We will perform a Dickey-Fuller test \cite{dickey1979distribution} on each element of a set of assets $\Omega$ with a significance level of $\alpha = 5\%$. We define a Python function \texttt{getRandomPerc($\Omega$)} that takes as input a set of crypto-assets $\Omega$ and returns the percentage of assets in that set that appear to follow a random walk, that is, for which we do not reject the null hypothesis "the time series is non-stationary". The result of \texttt{getRandomPerc($\Omega$)}, whose code is provided in Appendix \ref{appendix:getRandomPerc}, returns \texttt{69}\%. It seems that more than half of the cryptocurrencies follow a random walk.

There is often confusion between efficiency and random walk. Indeed, when reading the \href{https://fr.wikipedia.org/wiki/Hypoth\%C3\%A8se_des_march\%C3\%A9s_financiers_efficients}{Wikipedia} page on the efficient market hypothesis, one might think that an efficient market necessarily implies prices following a random walk. However, this is false. The market is not necessarily inefficient if prices do not follow a random walk because, as \cite{lardic2006efficience} states, "It suffices, for example, that the hypothesis of risk neutrality is not satisfied, or that individuals’ utility functions are not separable and additive \cite{leroy1982expectations}, meaning that it is impossible to separate consumption and investment decisions."

Many studies show that cryptocurrencies (most studies focus on Bitcoin) do not follow a random walk (\cite{palamalai2021testing}, \cite{aggarwal2019bitcoins}...). However, these studies mainly rely on the very restrictive assumption of autocorrelation, and conclude that the Bitcoin market is not efficient. Samuelson \cite{samuelson2016proof} already addressed this problem in his time and proposed a modification to the random walk hypothesis: the martingale model. 

This model is less restrictive than the random walk model because it imposes no condition on the autocorrelation of residuals. Very similar to the previous model, a price process $P_t$ follows a martingale if: $E[P_{t+1}|I_t] = P_t$, where $P_t$ is the price at time $t$ and $I_t$ is the information set at time $t$. Thus, under the martingale model, the current price is the sole (and best) predictor of the next price, even if there are successive dependencies in returns.

As previously noted, an analysis of most cryptocurrencies (the most widely used) shows that the returns of more than half of the assets seem to follow a random walk. With the martingale model, one might be tempted to assert that the crypto market is efficient.

However, many studies have investigated the relationship between Bitcoin and the martingale model (\cite{zargar2019informational}, \cite{nadarajah2017inefficiency}...) and conclude that the Bitcoin market is not efficient, mainly due to endogenous factors of an emerging and immature market, and the absence of traders relying on fundamental value.

It is difficult to extend this conclusion to the entire cryptocurrency market. However, we know that a study showing market inefficiency between 2012 and 2015 is not highly relevant for 2022, as much has happened since then (especially for Bitcoin).

Thus, we highlight the application of Lo's adaptive market hypothesis \cite{lo2004adaptive} to Bitcoin through a study \cite{khuntia2018adaptive}, which explains that efficiency improves over time. This study particularly well summarizes the evolution of crypto market returns: episodes of efficiency and inefficiency, creating opportunities for arbitrage and above-average returns, but an impossibility to predict these opportunities systematically or mathematically.

\subsubsection{Cryptocurrencies and Fundamental Value}

As explained by \cite{delcey2017efficient}, there are two definitions of an efficient market. Fama's definition implies that the randomness of a price is explained by the fact that prices converge toward the fundamental value. Samuelson's definition implies that unpredictable price variations are simply the result of competition among investors, regardless of fundamental value. This raises the following question: What is a fundamental value for a cryptocurrency?

According to \cite{biais2020equilibrium}, the fundamental value of Bitcoin (and by extension most other cryptocurrencies, as they hardly differ in their characteristics) lies in its stream of net transactional benefits, which depend on its future prices. These transactional benefits may, for instance, represent the ability to exchange money in an unstable economic and financial system (such as in Venezuela or Zimbabwe), or when exchanges are blocked or heavily taxed.

To determine the net value, \cite{biais2020equilibrium} consider various costs: limited convertibility, transaction fees from brokers, mining costs, and crash risk. They thus provide a definition of Bitcoin’s fundamental value (and technically of other cryptocurrencies) and answer the question of whether a cryptocurrency can have a fundamental value.

Obviously, this value differs depending on the cryptocurrency. For instance, if there is a strong demand for privacy in transactions, Monero (XMR) would dominate in volume, since it uses a private blockchain by default (making transactions untraceable, unlike Bitcoin where the blockchain is public and all transactions are identifiable).

However, the very idea that Bitcoin has a fundamental value is debated both in the media and academic literature. According to \cite{yermack2013bitcoin}, cryptocurrencies have no fundamental value because, if they did, there would be no incentive to mine cryptocurrency. According to \cite{hanley2013false}, Bitcoin's value merely floats relative to other currencies as a market estimate without any fundamental value to support it. \cite{woo2013bitcoin} suggests Bitcoin may have a certain fair value because of its features similar to fiat currencies (means of exchange and store of value), but without any other underlying basis. 

\cite{hayes2015decision} links the importance of Bitcoin’s mining network to the dependency of altcoin holders on Bitcoin, given that most altcoins must be exchanged into Bitcoin before being converted into fiat currency for real-world use. Furthermore, \cite{garcia2014digital} highlights the importance of mining production costs in the fundamental value of cryptocurrencies, as it provides a kind of “floor value”.

Cryptocurrencies are often criticized for being "backed by nothing", a misconception regarding the role of money in an economy. For example, according to the \href{https://www.federalreserve.gov/faqs/currency_12770.htm}{U.S. Federal Reserve}, 
``\textit{Federal Reserve notes are not redeemable in gold, silver, or any other commodity. Federal Reserve notes have not been redeemable in gold since January 30, 1934, when the Congress amended Section 16 of the Federal Reserve Act to read: "The said [Federal Reserve] notes shall be obligations of the United States….They shall be redeemed in lawful money on demand at the Treasury Department of the United States, in the city of Washington, District of Columbia, or at any Federal Reserve bank."}''

Beyond the purely economic definition of value (utility and scarcity), for which Bitcoin qualifies (its utility lying in being an alternative to the centralized financial system, and its scarcity from the 21 million unit limit and diminishing accessibility over time), there is also a subjective characteristic to this value. 

We highlight two relevant elements: network value and safe-haven value. According to Metcalfe’s law \cite{metcalfe1995metcalfe}, although nuanced \cite{odlyzko2005refutation}, the value of a network is proportional to the square of the number of its users: a single fax machine is useless, but the value of each fax increases with the total number of machines in the network. One could thus infer a similar characteristic for cryptocurrencies.

According to \cite{baur2010gold}, a safe-haven asset can be defined as one that is negatively correlated with equities during crises. Gold is often a reference point. Let us verify this. We cannot directly compare superimposed charts due to vastly different magnitudes:

\begin{figure}[H]
    \centering
   \includegraphics[width=1\linewidth]{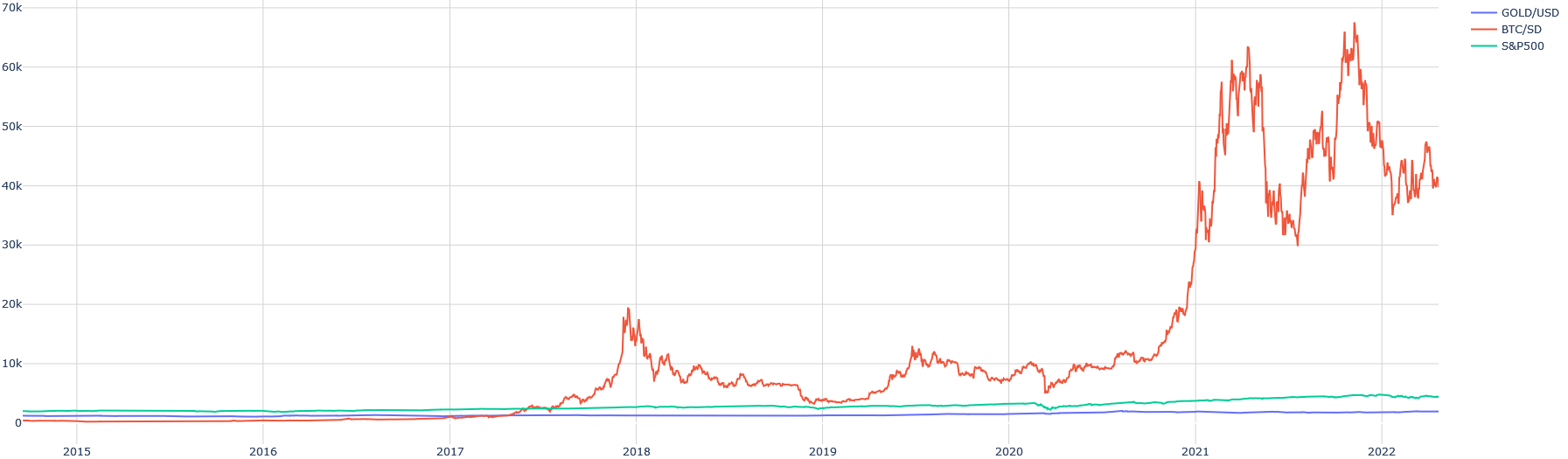}
    \caption{Correlation between BTC/USD, GOLD/USD, and S\&P500}
    \label{fig:13}
\end{figure}

Thus, we will separately analyze the correlation between S\&P500 crashes and BTC/USD prices:

\begin{figure}[H]
    \centering
   \includegraphics[width=1\linewidth]{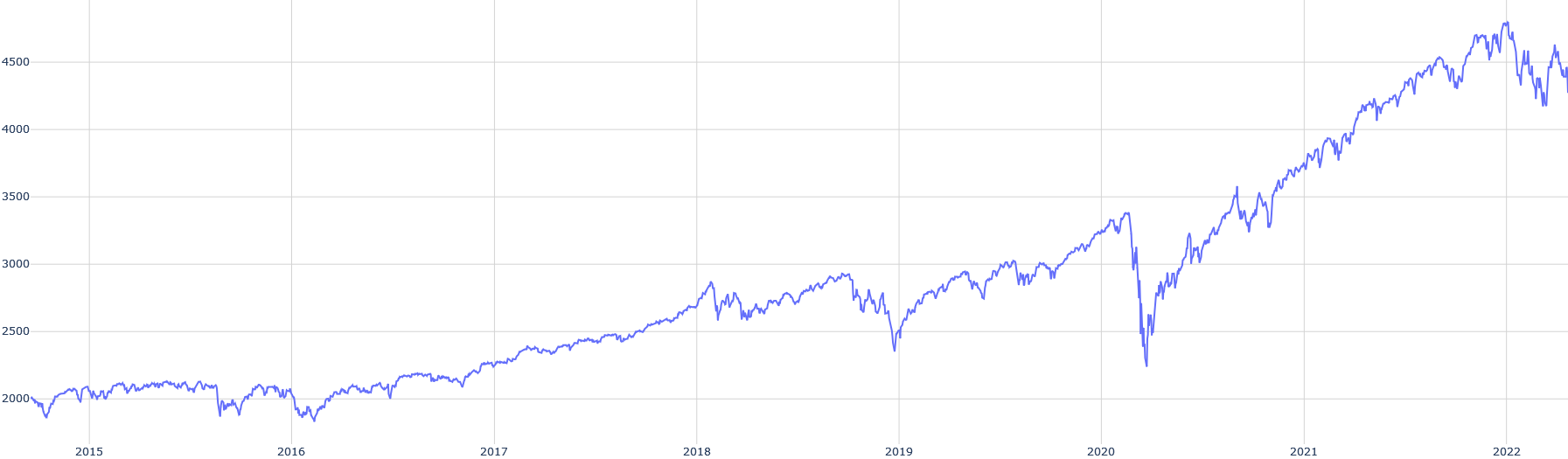}
    \caption{S\&P500 over the period available with BTC/USD}
    \label{fig:14}
\end{figure}

We notice graphical correlations during several crash periods:
\begin{itemize}
  \item Early 2018
  \item Late 2019
  \item Early 2020
  \item Early 2022
\end{itemize}

These correlations are weaker, or even negative, with gold:

\begin{figure}[H]
    \centering
   \includegraphics[width=1\linewidth]{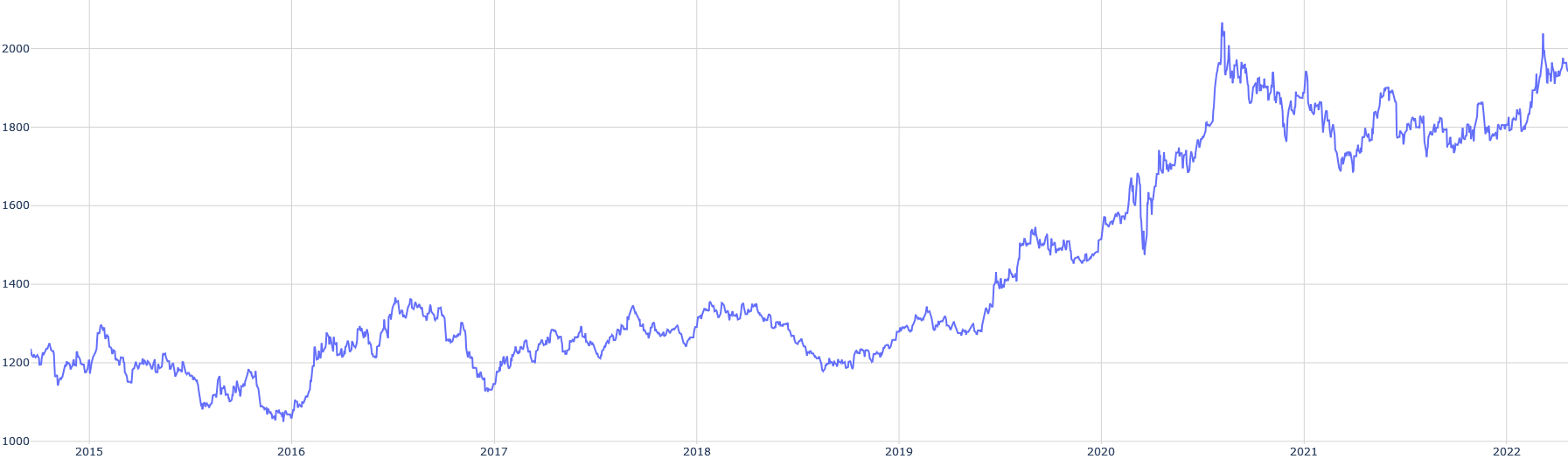}
    \caption{GOLD/USD over the period available with BTC/USD}
    \label{fig:15}
\end{figure}

Let us graphically check the correlation of daily returns:

\begin{figure}[H]
\centering
\begin{minipage}{.5\textwidth}
  \centering
  \includegraphics[width=1\linewidth]{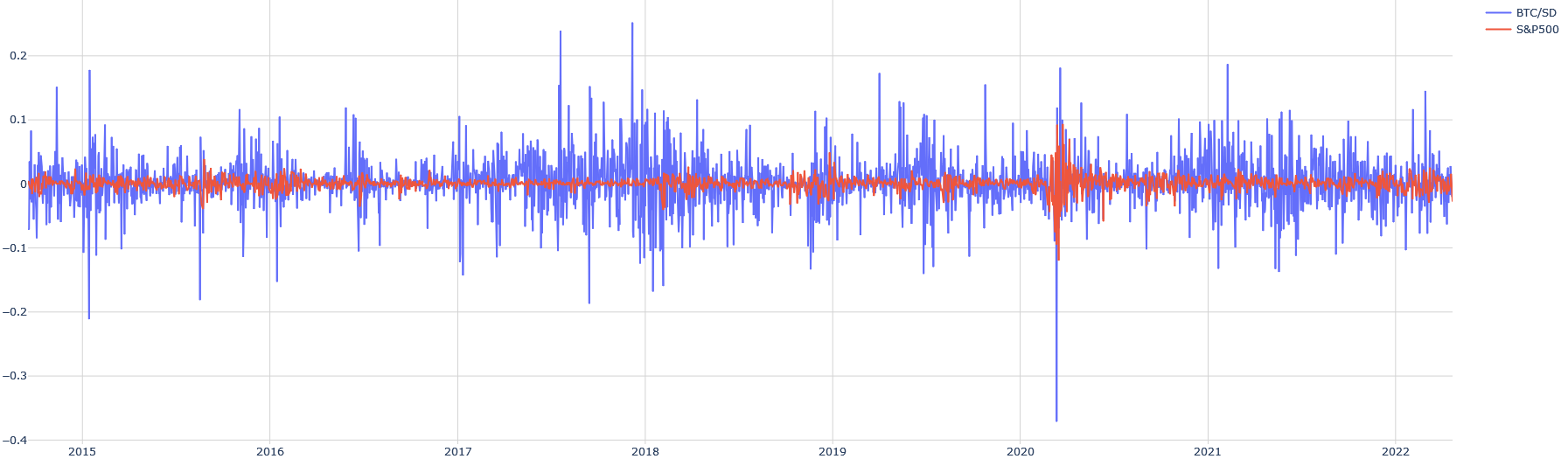}
  \caption{Correlation between BTC/USD and S\&P500 (daily returns)}
  \label{fig:16}
\end{minipage}%
\begin{minipage}{.5\textwidth}
  \centering
  \includegraphics[width=1\linewidth]{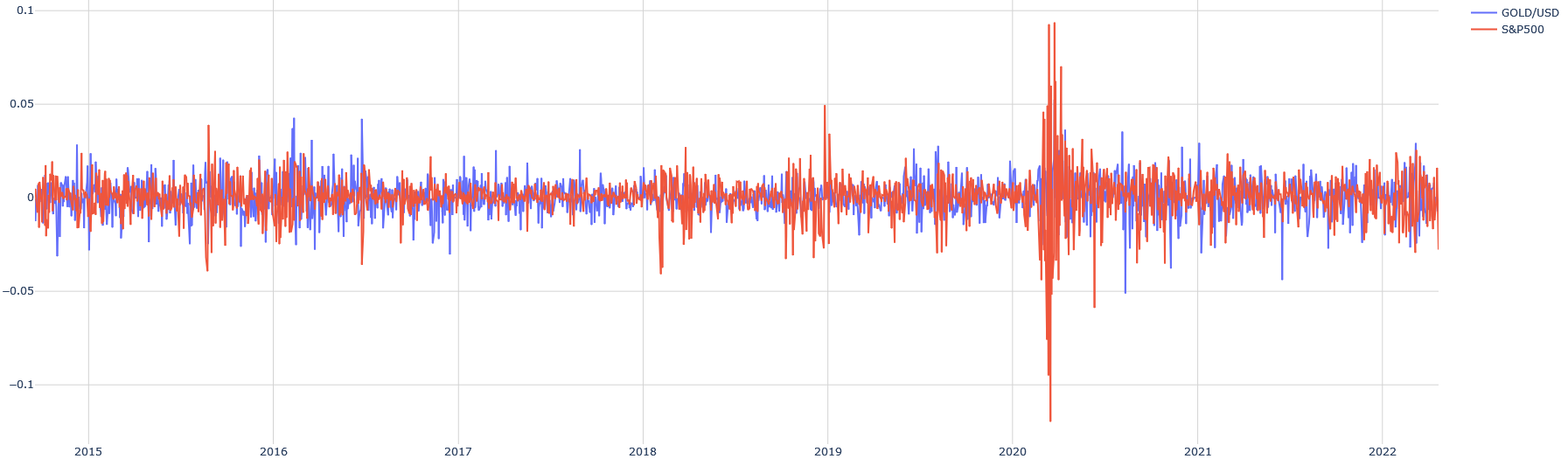}
  \caption{Correlation between GOLD/USD and S\&P500 (daily returns)}
  \label{fig:17}
\end{minipage}
\end{figure}

A numerical analysis of the correlation of daily returns over the entire period shows 16\% for Bitcoin with the S\&P500 and 5\% for gold with the S\&P500. Bitcoin does not appear to be a better safe-haven asset than gold, which is confirmed by other studies (\cite{smales2019bitcoin}, \cite{bouri2017hedge}).

\subsection{From Louis Bachelier to Contemporary Models}

Eugène Fama is not the inventor of the idea of a random market. We can trace it back to 1863, when Jules Regnault \cite{regnault1863calcul} proposed a model of randomly volatile markets. Then, in 1900, Bachelier \cite{bachelier1900theorie} formalized it. It was only from the 1930s that the random aspect of the market began to be considered, notably in the United States with the emergence of econometrics, and then, from the 1960s, financial economics in the United States started to connect the model to economic theory, giving rise to the theory of informational efficiency of financial markets. However, this theory, although constituting the foundation of the random walk model, would never achieve unanimous acceptance.
\\
In this subsection, we will present the theoretical models that have explained the variations of financial assets since 1900, from Louis Bachelier's theory of speculation to the present day.

\subsubsection{Modeling of Traditional Finance}

It is important to understand that the cryptocurrency market is not disconnected from traditional financial markets in its creation.

\begin{itemize}
  \item[$\blacksquare$] \textbf{The Louis Bachelier Model\\}
  Bachelier is a pioneer of modern finance in the sense that he was the first to use Brownian motion in modeling stock prices, five years before \cite{einstein1956investigations}. From his model, the Wiener process \cite{wiener1976collected} would later be formalized. The model simply explains that the stock market follows a Gaussian distribution. Of course, such a model today would not be considered rigorous, but for its time, it was already remarkably close to a correct model. Indeed, Brownian motion applied to stock price fluctuations is based on questionable assumptions: Markov chain (memoryless process), stationarity (constant mean and standard deviation), and normal distribution. We can clearly see, for example, for the four largest cryptocurrencies, that the distribution of daily returns is not really Gaussian:

\begin{figure}[H]
\centering
\begin{minipage}{.5\textwidth}
  \centering
  \includegraphics[width=1\linewidth]{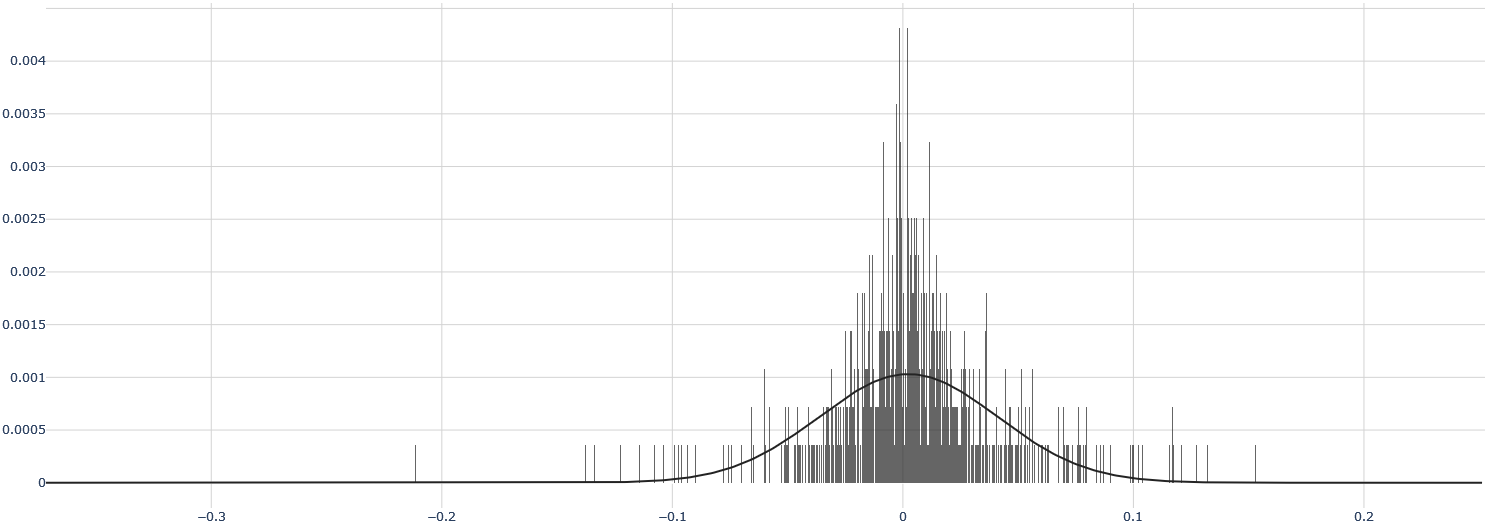}
  \caption{Distribution of daily returns for BTC/USD}
  \label{fig:18}
\end{minipage}%
\begin{minipage}{.5\textwidth}
  \centering
  \includegraphics[width=1\linewidth]{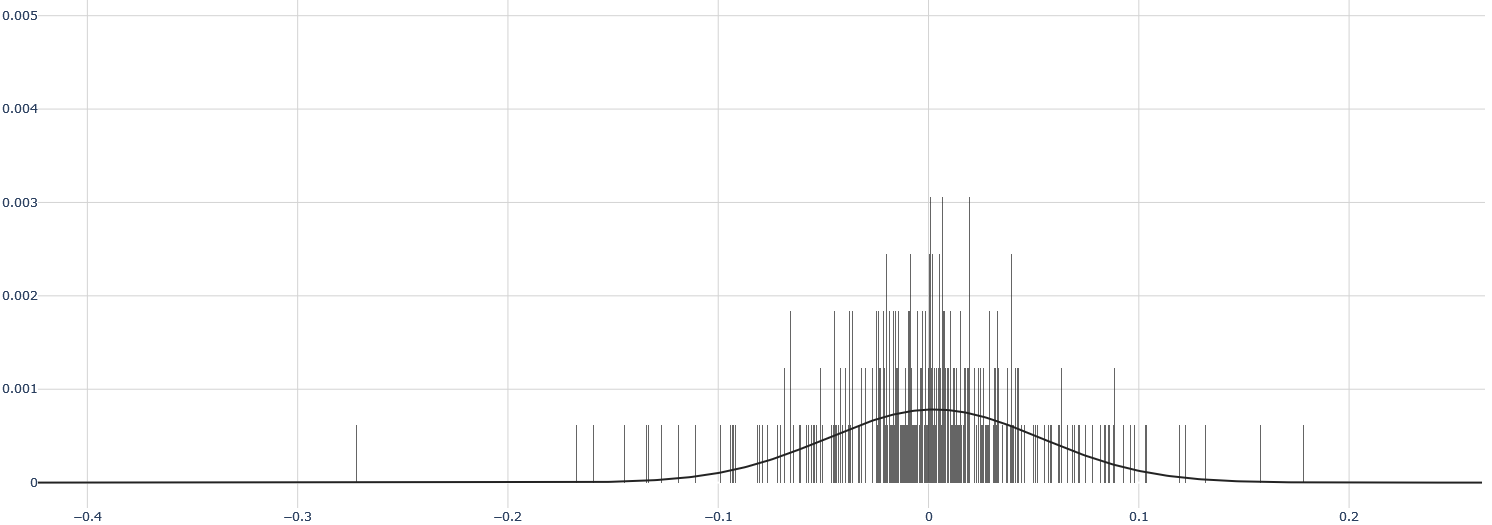}
  \caption{Distribution of daily returns for ETH/USD}
  \label{fig:19}
\end{minipage}
\end{figure}
\begin{figure}[H]
\centering
\begin{minipage}{.5\textwidth}
  \centering
  \includegraphics[width=1\linewidth]{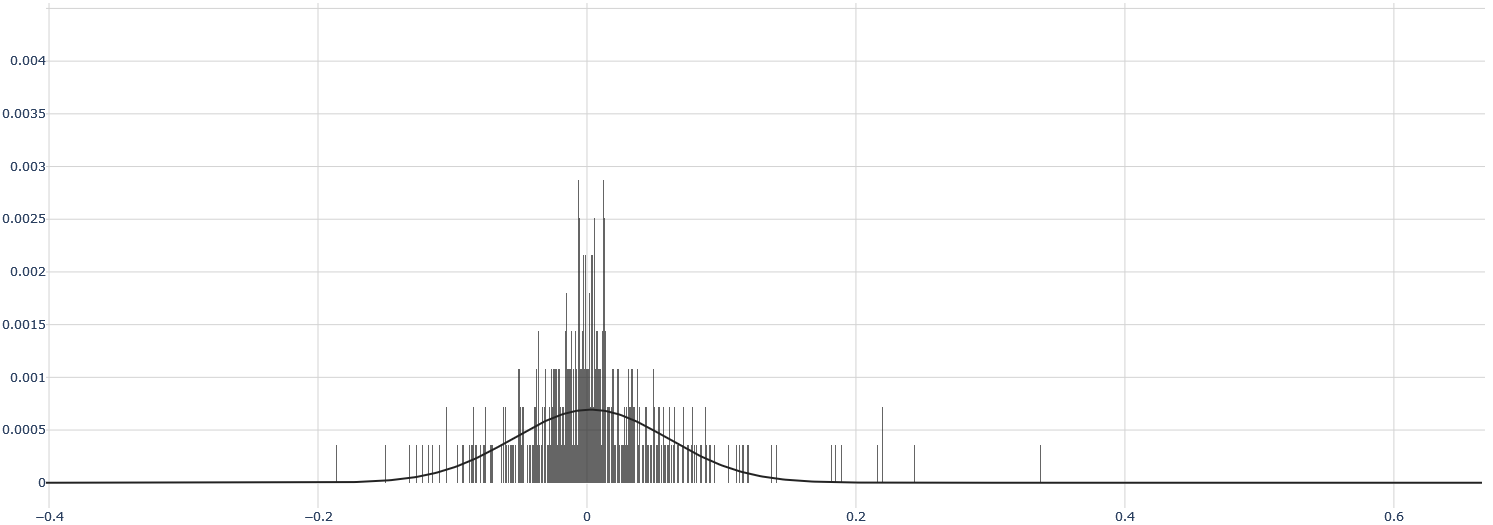}
  \caption{Distribution of daily returns for LTC/USD}
  \label{fig:20}
\end{minipage}%
\begin{minipage}{.5\textwidth}
  \centering
  \includegraphics[width=1\linewidth]{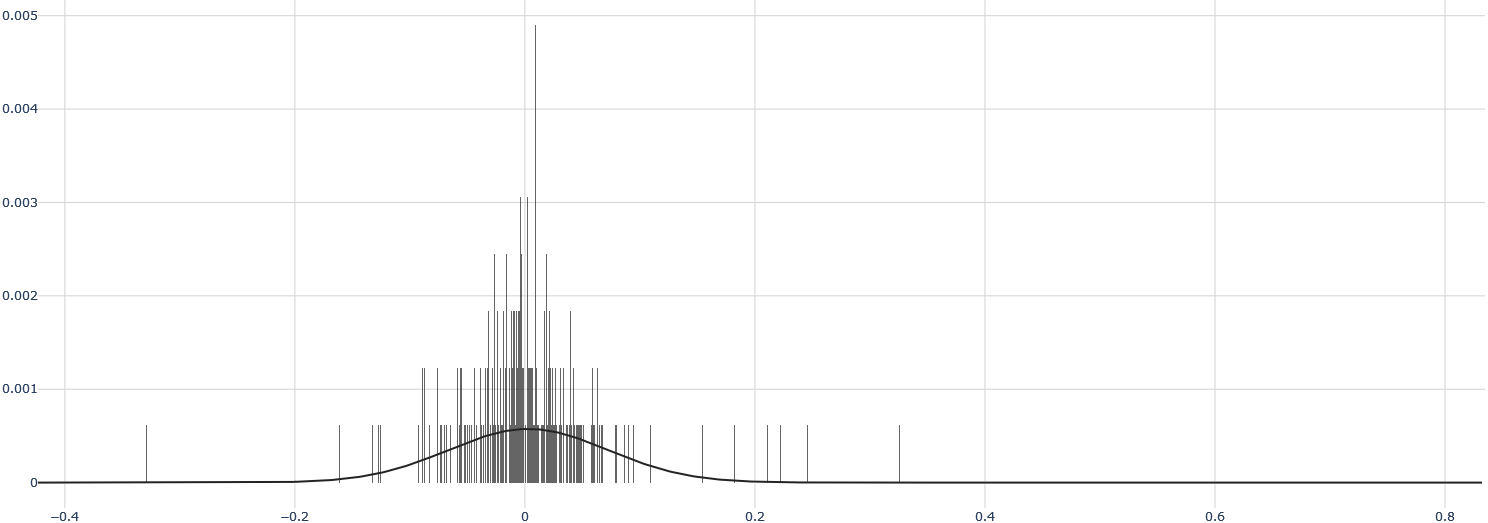}
  \caption{Distribution of daily returns for XRP/USD}
  \label{fig:21}
\end{minipage}
\end{figure}
It might be more appropriate to refer to a Lévy law or an $\alpha$-stable distribution.

\item[$\blacksquare$] \textbf{The Gordon-Shapiro Model\\}
The Gordon-Shapiro model \cite{gordon1956capital} is very well-known in finance and provides a very simple formula to model the price of a stock:
\begin{equation}
P_0 = \frac{D_1}{k-g}
\end{equation}
where $P_0$ is the theoretical value of the stock, $D_1$ the anticipated dividend for the first period, $k$ the expected return rate for the shareholder, and $g$ the growth rate of the gross earnings per share.\\
The first thing to note is that this model is useless for the crypto market: there are no dividends. Therefore, this model can be dismissed, even though it is attractive.

\item[$\blacksquare$] \textbf{Contemporary Models\\}
With the development of quantitative finance and derivative pricing, many models have emerged, one of the most famous being the Black \& Scholes model \cite{black2019pricing}. However, as with the binomial model (Cox, Ross \& Rubinstein model), the problem of constant volatility of the underlying assets appeared. Indeed, in the Black \& Scholes formula:
\begin{equation}
    C = S_tN(d_1)-Ke^{-rt}N(d_2)
\end{equation}
where:
\begin{equation}
    d_1 = \frac{\ln\frac{S_t}{K} + (r+\frac{\sigma^2}{2})t}{\sigma \sqrt{t}}
\end{equation}
\begin{equation}
    d_2 = d_1 - \sigma \sqrt{t}
\end{equation}
with:\\
$C$ the price of the call option\\
$P$ the stock price\\
$K$ the strike price\\
$r$ the risk-free interest rate\\
$t$ the time in years to maturity\\
$N$ a normal distribution\\
$\sigma$ the volatility of the underlying asset\\
We notice that volatility is considered constant. This led to the development of stochastic volatility models, treating the volatility of the underlying as a random process. As explained, for instance, by \cite{mantegna1999introduction}, the price of an asset can be characterized by a standard geometric Brownian motion:
\begin{equation}
    dS_t = \mu S_tdt + \sigma S_tdW_t
\end{equation}
with:\\
$\mu$ the drift (often negligible)\\
$\sigma$ constant volatility\\
$dW_t \hookrightarrow N(0,1)$ an increment of Brownian motion\\
then replacing $\sigma$ by a process $\nu_t$. This is indeed how the Heston model \cite{heston1993closed} is built, one of the most well-known stochastic volatility models. Its formulas are:
\begin{equation}
    dS_t = rS_tdt+\sqrt{V_t}S_tdW_{1t}
\end{equation}
with $V_t$ the instantaneous variance:
\begin{equation}
    dV_t = \kappa (\theta - V_t)dt + \sigma \sqrt{V_t}dW_{2t}
\end{equation}
where:\\
$S_t$ the asset price at time $t$\\
$r$ the risk-free interest rate\\
$\sqrt{V_t}$ the volatility (standard deviation) of the price\\
$\sigma$ the volatility of the volatility (i.e., of $\sqrt{V_t}$)\\
$\theta$ the long-term variance\\
$\kappa$ the reversion rate to $\theta$\\
$dt$ an infinitesimally small time increment\\
$W_{1t}$ the Brownian motion for the asset price\\
$W_{2t}$ the Brownian motion for the variance of the asset price\\
with the property that, for Brownian motions, $W_0 = 0$, the $W_t$ are independent, and $W_t$ is continuous in $t$.\\
This model seems well suited for modeling the price of cryptocurrencies. Indeed, \cite{kachnowski2020empirical} explains that an adaptation of the Heston model to Bitcoin improves the accuracy of predictions over time windows ranging from 7 days to 2 months. However, as shown by \cite{gatheral2018volatility}, the log-volatility is not actually a classic Brownian motion but rather a fractional Brownian motion, as in the \textit{Fractional Stochastic Volatility Model} by \cite{comte2012affine}, but with a Hurst exponent of 0.1 (and not 0.5 as in \cite{comte2012affine}, who did not take into account the \textit{rough} aspect of volatility).
\end{itemize}

\subsubsection{Modeling Crypto-Finance}

\begin{itemize}
  \item[$\blacksquare$] \textbf{Quantitative Theory of (Crypto)Currency\\}
As we know \cite{fisher2006purchasing},
\begin{equation}
    MV=PY
\end{equation}
where:\\
$M$ is the money supply\\
$V$ is the velocity of money\\
$P$ is the price level\\
$Y$ is the output of the economy\\
Let's adapt this model to cryptocurrencies.\\
For $M$, it is simple: it is constant at 21 million. However, we can already anticipate that $M$ tends towards 0. Indeed, 21 million is the maximum number of Bitcoins that can be mined. Once mined, Bitcoins can disappear for several reasons: lost passwords, hacking, computer errors, etc. For $V$, it is more complicated. We would need to differentiate between economically meaningful transactions and meaningless ones. And this is very difficult, even though all transactions are listed on the Blockchain, the reasons behind them are not. Thus, we cannot distinguish "real" transactions from "fake" ones.\\
For $P$, it refers to the goods and services that can be purchased with Bitcoin. In November 2020, the Venezuelan branch of Pizza Hut accepted Bitcoin. On that day, you could buy around 1,800 pizzas (worth approximately 10 USD each) with one Bitcoin. Today, you could buy around 4,000 pizzas with one Bitcoin. Thus, $P$ has been continuously falling for BTC/USD. For $Y$, it represents the amount of goods and services available for purchase and sale. We can admit that very few goods and services are currently bought and sold with cryptocurrencies. Thus, over time, cryptocurrencies are expected to depreciate. Indeed, we know that the number of Bitcoins in circulation initially increases (then will decrease), which should induce inflation. However, the opposite is observed. If $Y$ is exogenous to Bitcoin (goods and services offered are not really dependent on Bitcoin's price), and $M$ is constant, then $V$ will influence $P$. In this case, two scenarios arise: if Bitcoin (same reasoning for other cryptocurrencies) is merely a means of exchange without any fundamental value, $V$ will increase, as it will become just another payment option for households. If Bitcoin is rather seen as a store of value, with a fundamental value, then households will invest and hold their Bitcoins, causing $V$ to decrease, which will raise $P$. Studies, including \cite{pernice2020cryptocurrencies}, show a link between price and velocity in cryptocurrencies.

  \item[$\blacksquare$] \textbf{Other Models of the Crypto Market\\}
  \cite{cretarola2018modeling} propose modeling the crypto market by the interest it generates. They explore the link between Bitcoin's price behavior and investor attention in the network. They conclude that the attention index impacts Bitcoin's price through dependence of the drift and diffusion coefficients and potential correlation between the sources of randomness represented by Brownian motions. \cite{hou2020pricing} propose a model for pricing crypto options, SVCJ (\textit{stochastic volatility with a correlated jump}), similar to \cite{pascuccistochastic}, and compare it with the \textit{cojump} model of \cite{bandi2016price}. It is very likely that the future of cryptocurrency price modeling will develop towards derivative products.
\end{itemize}

\subsection{Time Series Studies and Analyses}
We are still considering the case where the crypto market is efficient. Thus, it is impossible to predict its price movements, regardless of the methods employed. However, these methods are still widely used by both retail and professional investors. Therefore, we will examine these methods to understand whether they can be effective in prediction. Nevertheless, we will see that it is sometimes difficult to answer this question with a simple yes or no.

\subsubsection{Fundamental Analysis}
To perform fundamental analysis on a company, there are a number of well-established methods (financial ratios, EBITDA, cash flows, etc.). For cryptocurrencies, there are not really established methods. We have therefore chosen 5 themes. We cannot develop a full analysis due to the lack of data, but these indicators can, in our opinion, allow a good fundamental analysis:
\begin{itemize}
    \item[$\blacksquare$] Supply Measures:
    \begin{itemize}
        \item Is the number of coins fixed in advance?
        \item How many coins have been mined, and how many remain?
        \item What is the inflation rate?
        \item What is the coin-to-flow ratio?
        \item What is the granularity of the coins?
    \end{itemize}
    \item[$\blacksquare$] Value Measures:
    \begin{itemize}
        \item What is the current price?
        \item What is the current gross market capitalization?
        \item What is the current net market capitalization (excluding lost coins)?
        \item What is the interest rate (borrowing cryptocurrencies)?
        \item What are the yearly high and low points?
        \item What are the returns by day, week, month, year, and overall?
    \end{itemize}
    \item[$\blacksquare$] Network Activity Measures:
    \begin{itemize}
        \item How many active addresses are there?
        \item How many new addresses are there?
        \item How many transactions are there?
        \item What is the average transaction size?
    \end{itemize}
    \item[$\blacksquare$] Broker Activity Measures:
    \begin{itemize}
        \item What is the total traded volume?
        \item On how many brokers is the cryptocurrency listed?
        \item What is the broker flow?
        \item In which geographical areas do the flows occur?
    \end{itemize}
    \item[$\blacksquare$] Mining Measures:
    \begin{itemize}
        \item What is the consensus mechanism (Proof of Work, Proof of Stake, etc.)?
        \item What is the governance of the mining network?
        \item How long does it take to mine a block?
        \item How are miners rewarded?
        \item What are the median fees?
        \item What is the hash rate?
    \end{itemize}
\end{itemize}

\subsubsection{Chartist / Technical Analysis}
Technical analysis aims to predict a price using future prices, and more precisely through repetitive patterns or technical indicators. Beyond the SMA tested previously, let's simply perform a graphical analysis. Let's take the RSI on BTC/USD:
\begin{figure}[H]
    \centering
   \includegraphics[width=1\linewidth]{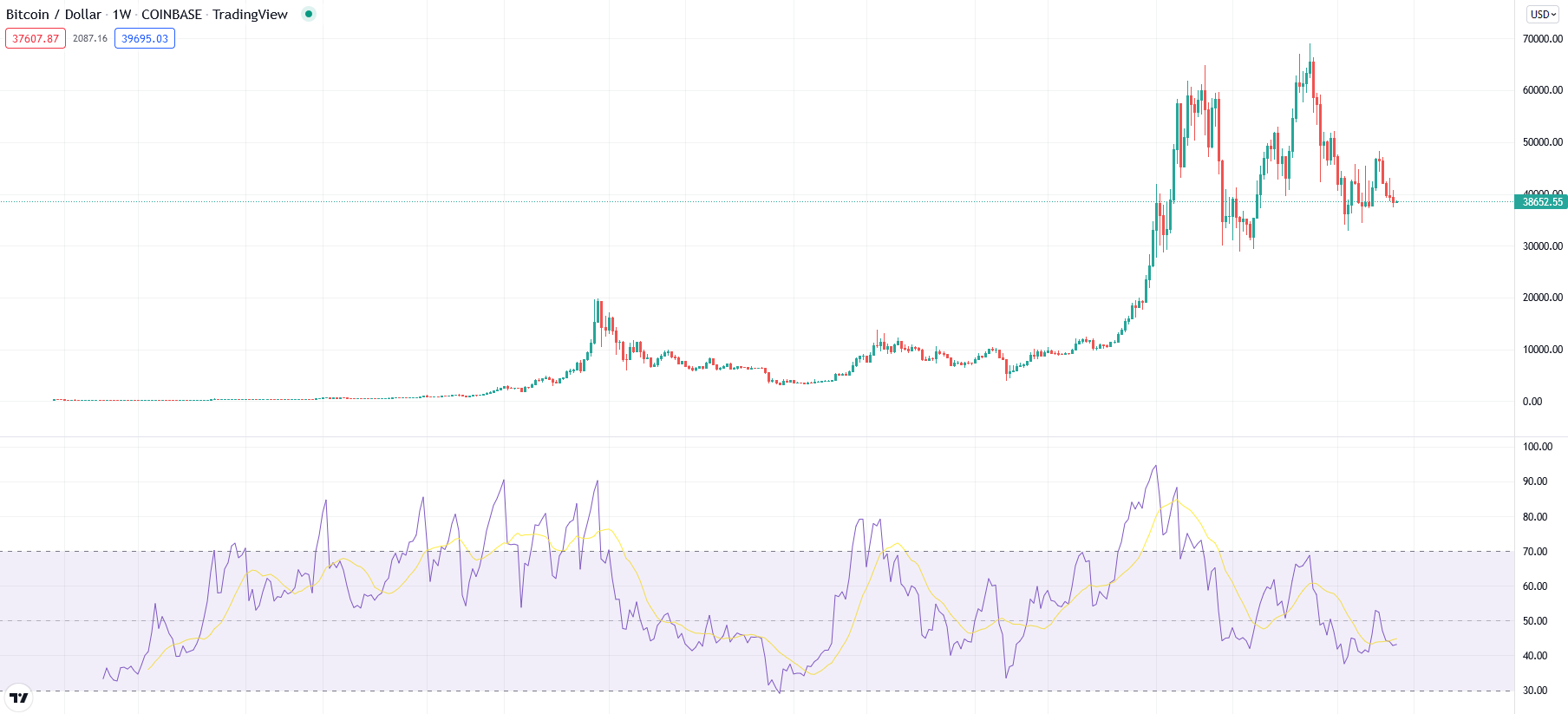}
    \caption{RSI Signals for BTC/USD}
    \label{fig:22}
\end{figure}
This indicator tells us that when it is below 30, we should buy, and above 70, we should sell. It is clearly seen that the RSI is useless for a long-term vision: what is the use of selling at 10,000 in 2018 when one could simply buy, hold, and sell at 60,000 in 2022? Now, let's take another very famous technical indicator: the SAR.
\begin{figure}[H]
    \centering
   \includegraphics[width=1\linewidth]{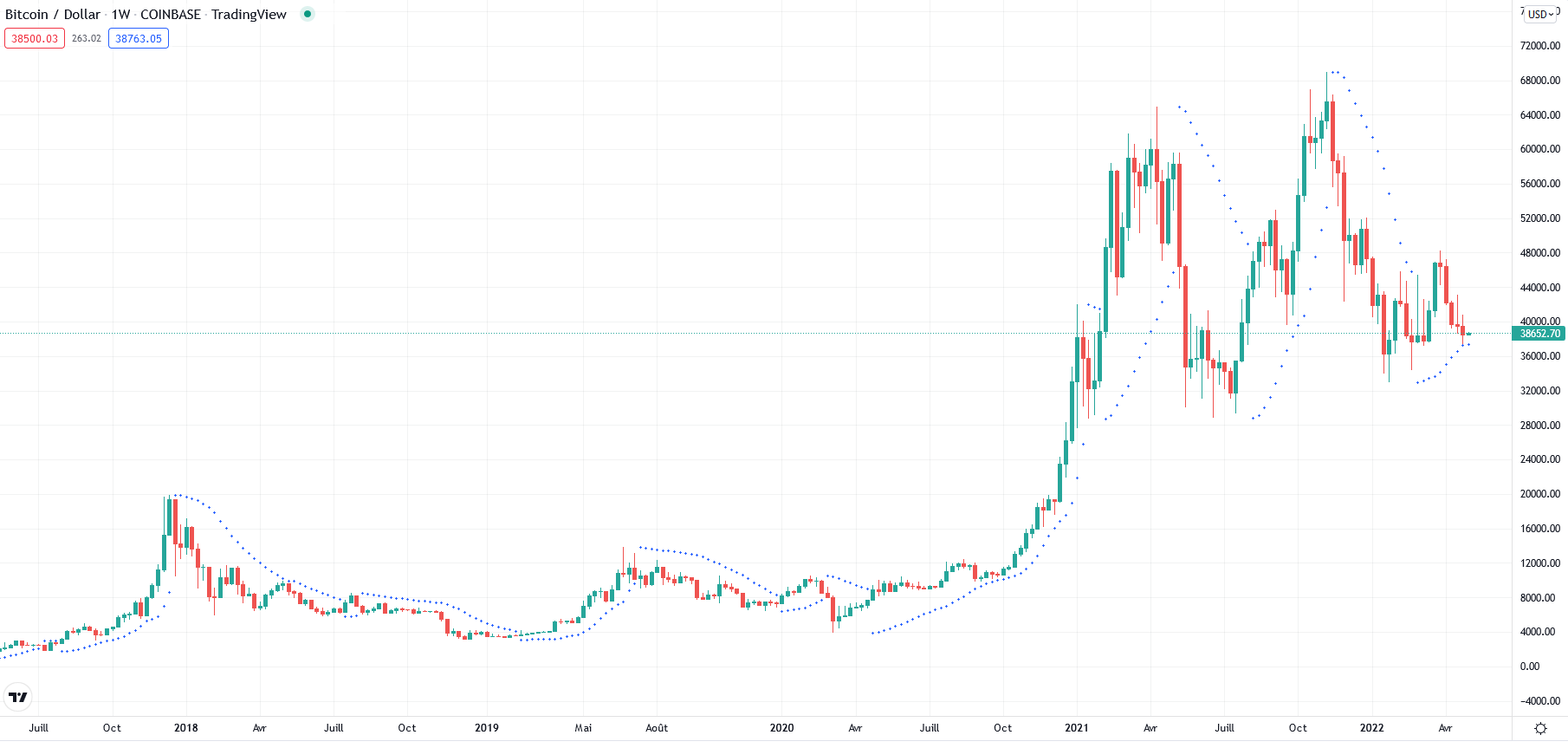}
    \caption{SAR Signals for BTC/USD}
    \label{fig:23}
\end{figure}

This indicator tells us that when the blue dots are below the candlesticks, we should buy, and when they are above, we should sell (first appearance of dots per sequence). The same reasoning applies: what is the point of selling in 2018? These indicators are ultimately only signals for day-trading, with the aim of making quick profits. However, statistics show that more than 70\% of day-traders lose money. Yet, they all have access to all available indicators. Technical analysis would therefore seem useless both in the long term and in the short term, a priori. According to \cite{park2007we}, the literature on the subject is inconclusive: some studies are positive, others negative, and others mixed.

\subsubsection{Machine Learning}
Let's now check the effectiveness of Machine Learning in predicting cryptocurrency prices. We will not test all algorithms, but only two. The first, Support Vector Machine classification, was introduced by \cite{cortes1995support}. It consists of classifying "good" trades from "bad" trades. For this, we create a Python function \texttt{getAverageAccuracy($\Omega,n$)}, which takes as parameters $\Omega$ and $n$ the window for technical indicators and returns the average accuracy percentage of our model across all tested cryptocurrencies (over 100). The features considered are: price (OHLC), previous prices, previous returns, SAR, RSI, SMA, ADX, ATR, and 80\% training dataset. The function, whose code is in Appendix \ref{appendix:getAverageAccuracy}, returns \texttt{38\%}. This is low. Here are the confusion matrices for the 4 largest cryptocurrencies (read as "Perfect prediction on the top-left/bottom-right diagonal, inverse prediction on the bottom-left/top-right"):

\begin{figure}[H]
\centering
\begin{minipage}{.5\textwidth}
  \centering
  \includegraphics[width=0.9\linewidth]{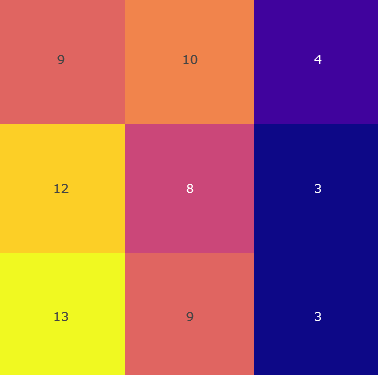}
  \caption{Confusion matrix for BTC/USD}
  \label{fig:24}
\end{minipage}%
\begin{minipage}{.5\textwidth}
  \centering
  \includegraphics[width=0.9\linewidth]{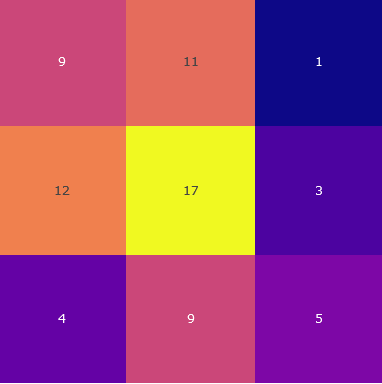}
  \caption{Confusion matrix for ETH/USD}
  \label{fig:25}
\end{minipage}
\end{figure}
\begin{figure}[H]
\centering
\begin{minipage}{.5\textwidth}
  \centering
  \includegraphics[width=0.9\linewidth]{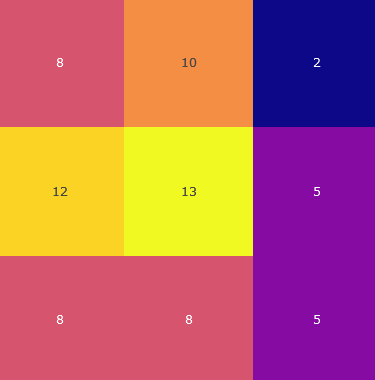}
  \caption{Confusion matrix for LTC/USD}
  \label{fig:26}
\end{minipage}%
\begin{minipage}{.5\textwidth}
  \centering
  \includegraphics[width=0.9\linewidth]{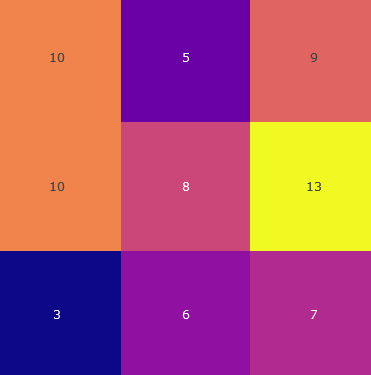}
  \caption{Confusion matrix for XRP/USD}
  \label{fig:27}
\end{minipage}
\end{figure}

The second model is ARIMA, introduced by \cite{box2015time}, and it aims to predict future trends. The results of our model are as follows:

\begin{figure}[H]
    \centering
   \includegraphics[width=1\linewidth]{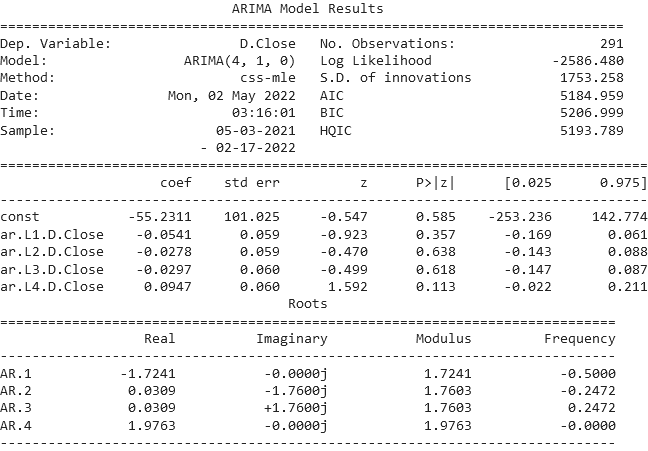}
    \caption{Results of the ARIMA model}
    \label{fig:28}
\end{figure}

\section{The Cryptocurrency Market is Inefficient}
In 1981, Robert Shiller \cite{shiller1980stock} showed a higher volatility than that predicted by the rational behavior of agents. Shiller concluded that no rationality could explain the observed volatility, which ultimately had no link with dividend expectations. Thus, if the market is inefficient, it is possible to achieve performances superior to the market.

\subsection{Robert Shiller and the Notion of an Inefficient Market in Terms of Arbitrage}
This section deals with elements that prove that the Bitcoin market admits arbitrage opportunities. For example, we observe that the price of Bitcoin varies from one exchange to another. This is even more true for the altcoin market. Intuitively, we can imagine that the price will tend to move closer to the average price across exchanges.

\subsubsection{Volatility and Expected Dividends}
In his book, \cite{shiller2015irrational}, Shiller shows the difference between stock price volatility and expected dividends:

\begin{figure}[H]
    \centering
   \includegraphics[width=1\linewidth]{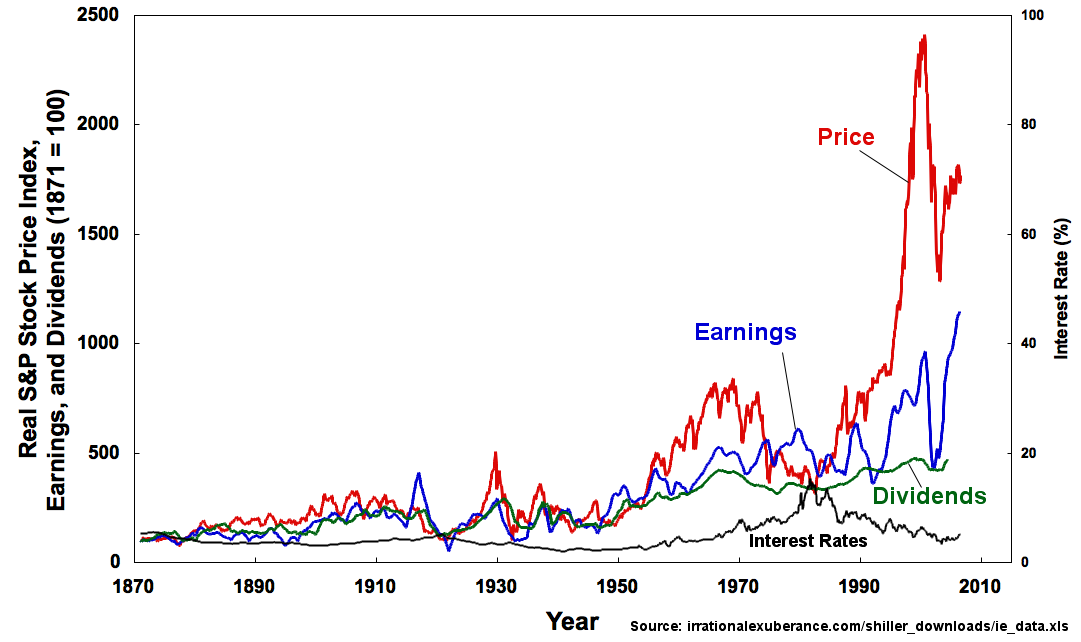}
    \caption{Evolution of the S\&P500 and dividends}
    \label{fig:29}
\end{figure}

According to him: "the price-to-earnings ratio is still (as of 2005) far from its historical average from the mid-20th century. Investors place too much trust in the market and overestimate the positive developments of their investments without sufficiently hedging against a market downturn." It is therefore difficult to determine whether the crypto market is inefficient based solely on this information, as cryptocurrencies do not pay dividends.

\subsubsection{Behavioral Finance and Market Anomalies}
Shiller introduces the concept of behavioral finance. In the crypto market, we mainly think of herd behavior: investors buy simply because other investors are buying. This phenomenon is less visible in day-trading because time scales are too short to draw conclusions about the trend. Indeed, the 2017 bubble still took some time to form and partially burst.

\subsubsection{Speculative Bubbles}
When it comes to cryptocurrencies, speculative bubbles are often mentioned. It is true that cryptocurrencies provide fertile ground for such phenomena, but this only matters for medium-term investors. A long-term investor will mainly seek to minimize diversifiable risk through cryptocurrencies, while a short-term trader will hope to enter the market before a hype event. Moreover, these hypes can sometimes be artificially created by one or several people, sometimes even behind fraudulent projects. Over time, as projects repeat, fraud risks decrease, and hypes also tend to diminish, making the crypto market increasingly efficient and reducing the possibility of bubbles.

\subsection{Informational Inefficiency}
We will look at scenarios where information asymmetries allow an individual or a group to achieve superior returns to the market. In such situations, cryptocurrency prices do not reflect all available information.

\subsubsection{Market Manipulation}
The most famous example is the public use of Twitter by Elon Musk, with each of his crypto-related tweets causing abrupt movements in the crypto market. By deduction, we can imagine similar scenarios involving other public figures, broker managers, intermediaries, etc.

\subsubsection{Pump \& Dump}
Pump \& Dump was a strong practice during the early days of crypto hype. It consisted of gathering the largest possible group of users around a well-promoted cryptocurrency. The initiator of the movement would encourage the entire community to engage with the project for a single purpose: to artificially inflate the price of the cryptocurrency. Once the cryptocurrency reached a satisfactory price, the initiator—who had taken care to invest as much as possible when the crypto was worth nothing—would sell everything and exit the project. This type of phenomenon was also seen with ICOs.

\subsubsection{Natural Language Processing}
Natural Language Processing (NLP) can be used to analyze market sentiment without manually reading content. For example, the bot, whose code is in Appendix \ref{appendix:nlp}, returns the following results:

\begin{figure}[H]
    \centering
   \includegraphics[width=1\linewidth]{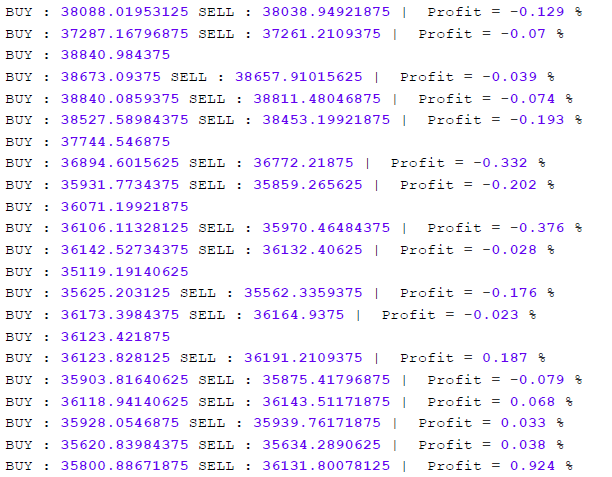}
    \caption{Results from the NLP trading bot}
    \label{fig:29}
\end{figure}

\subsection{Operational Inefficiency}
The price of Bitcoin can be predicted if one knows in advance the factors likely to influence the network as a whole, or a significant part of it. We will explore whether one or several elements hindering cryptocurrency exchanges can induce market movements.

\subsubsection{At the Macroscopic Scale}
We can take the example of countries that ban cryptocurrencies. These bans have a notable effect on liquidity, or cases of massive adoption like in El Salvador or the Marshall Islands, or the future rise of CBDCs (such as the digital euro). Environmental concerns, which are becoming a major issue, also hinder liquidity, as cryptocurrencies require a significant amount of electricity resources.

\subsubsection{At the Mesoscopic Scale}
Certain cryptocurrencies can have a negative impact on others. For example, Monero, with its private blockchain, can very well absorb all the demand for cryptocurrencies that also aim to respect user privacy. The same goes for issues related to transaction speed.

\subsubsection{At the Microscopic Scale}
There are usage barriers to crypto-assets among households that strongly impact the markets, such as the prohibition of cryptocurrency usage for minors, broker restrictions regarding certain trading positions, security risks and broker compliance concerning suspicious activities, tainted bitcoins, and money laundering (KYC/AML requirements), the impact of taxation on crypto-related capital gains, and various hacks. It can be observed that these phenomena have much less impact on the markets than macro or even meso factors. However, households and discretionary traders still represent a large part of the crypto market landscape.

\section{Conclusion}
In conclusion, by default, it is not possible to predict Bitcoin since it is an asset very similar in nature to others (notably the stock market), but, as with any market, there are moments when the market is inefficient, and thus it is possible to profit from these moments and predict Bitcoin prices accurately.

Among the limitations, we focus only on the spot market, we do not consider the influence of other cryptocurrencies, and we are limited in our expertise in time series analysis.

Among public policy recommendations, we agree with the view of \cite{brito2014bitcoin} regarding the regulation of brokers and particularly of derivatives products, which are becoming increasingly significant in the crypto market.

\appendix
\section{\texttt{isRandomBetter($\Omega, n, k$)}}
\label{appendix:isRandomBetter}
\begin{lstlisting}[language=Python]
# The set Omega is a subset of all cryptocurrencies on the market (between 10,000 and 20,000)
Omega = ['1INCH-USD', 'AAVE-USD', 'ACH-USD', 'ADA-USD', 'AERGO-USD', 'AGLD-USD',
         'AIOZ-USD', 'ALCX-USD', 'ALGO-USD', 'ALICE-USD', 'AMP-USD', 'ANKR-USD',
         'APE-USD', 'API3-USD', 'ARPA-USD', 'ASM-USD', 'ATOM-USD', 'AUCTION-USD',
         'AVAX-USD', 'AVT-USD', 'AXS-USD', 'BADGER-USD', 'BAL-USD', 'BAND-USD', 
         'BAT-USD', 'BCH-USD', 'BICO-USD', 'BLZ-USD', 'BNT-USD', 'BOND-USD', 
         'BTC-USD', 'BTRST-USD', 'CHZ-USD', 'CLV-USD', 'COMP-USD', 
         'COTI-USD', 'COVAL-USD', 'CRO-USD', 'CRPT-USD', 'CRV-USD', 'CTSI-USD', 
         'CTX-USD', 'CVC-USD', 'DAI-USD', 'DASH-USD', 'DDX-USD', 'DESO-USD', 
         'DIA-USD', 'DNT-USD', 'DOGE-USD', 'DOT-USD', 'ENJ-USD', 'ENS-USD', 
         'EOS-USD', 'ERN-USD', 'ETC-USD', 'ETH-USD', 'FARM-USD', 
         'FET-USD', 'FIDA-USD', 'FIL-USD', 'FORTH-USD', 'FOX-USD', 'FX-USD', 
         'GALA-USD', 'GFI-USD', 'GLM-USD', 'GNT-USD', 'GODS-USD', 
         'GRT-USD', 'GTC-USD', 'GYEN-USD', 'HIGH-USD', 'ICP-USD', 'IDEX-USD', 
         'IMX-USD', 'INV-USD', 'IOTX-USD', 'JASMY-USD', 'KEEP-USD', 'KNC-USD', 
         'KRL-USD', 'LCX-USD', 'LINK-USD', 'LOOM-USD', 'LPT-USD', 'LQTY-USD', 
         'LRC-USD', 'LTC-USD', 'MANA-USD', 'MASK-USD', 'MATIC-USD', 'MCO2-USD', 
         'MDT-USD', 'MINA-USD', 'MIR-USD', 'MKR-USD', 'MLN-USD', 'MPL-USD', 
         'MUSD-USD', 'NCT-USD', 'NKN-USD', 'NMR-USD', 'NU-USD', 'OGN-USD', 
         'OMG-USD', 'ORCA-USD', 'ORN-USD', 'OXT-USD', 'PERP-USD', 
         'PLA-USD', 'PLU-USD', 'POLS-USD', 'POLY-USD', 'POWR-USD', 'PRO-USD', 
         'QNT-USD', 'QSP-USD', 'QUICK-USD', 'RAD-USD', 'RAI-USD', 'RARI-USD', 
         'RBN-USD', 'REN-USD', 'REP-USD', 'REQ-USD', 'RGT-USD', 'RLC-USD', 
         'RLY-USD', 'RNDR-USD', 'SHIB-USD', 'SHPING-USD', 'SKL-USD', 'SNT-USD', 
         'SNX-USD', 'SOL-USD', 'SPELL-USD', 'STORJ-USD', 'STX-USD', 'SUKU-USD', 
         'SUPER-USD', 'SUSHI-USD', 'SYN-USD', 'TBTC-USD', 'TRAC-USD', 'TRB-USD', 
         'TRIBE-USD', 'TRU-USD', 'UMA-USD', 'UNFI-USD', 'UNI-USD', 'UPI-USD', 
         'USDC-USD', 'USDT-USD', 'UST-USD', 'VGX-USD', 'WBTC-USD', 
         'WCFG-USD', 'WLUNA-USD', 'XLM-USD', 'XRP-USD', 'XTZ-USD', 'XYO-USD', 
         'YFI-USD', 'YFII-USD', 'ZEC-USD', 'ZEN-USD', 'ZRX-USD']
\end{lstlisting}

\begin{lstlisting}[language=Python]
# This function returns a list of returns for each asset
def loadChanges(Omega):
    changes = []
    for asset in Omega:
        # Import time
        time.sleep(1)
        # We use the yFinance library to get the data
        df = yf.download(asset, period = '2y', interval = '1d', progress=False)
        if df.empty:
            continue
        oneYear= df.loc['2021-01-01':'2022-01-01']
        if oneYear.empty or len(df) < 360:
            continue
        # Import pandas
        s = pd.Series(list(oneYear['Close']))
        if not s[s.isin([0])].empty:
            continue
        else:
            start = oneYear.iloc[0]['Close']
            final = oneYear.iloc[-1]['Close']
            change = ((final-start)/start)*100
            changes.append(change)    
    print("len Valid assets : ", len(changes), " (only consider the 1st print!)")
    return changes
\end{lstlisting}

\begin{lstlisting}[language=Python]
# This function returns the average of returns
def getMeanChanges(Omega):
    changes = loadChanges(Omega)
    return sum(changes)/len(changes)
\end{lstlisting}

\begin{lstlisting}[language=Python]
# This function randomly selects a portfolio of crypto-assets among those available in Omega
def generateRandomPortfolio(Omega, k):
    randomPortfolio = []
    for _ in range(k):
        # Import random
        randomPortfolio.append(random.choice(Omega))
    return randomPortfolio
\end{lstlisting}

\begin{lstlisting}[language=Python]
# This function returns the percentage of portfolios with an average return higher than the average return 
# of the assets in Omega
def getPercentageHigherThanAverage(Omega, NbIter, k):
    nbHigher = 0
    averageReturns = getMeanChanges(Omega)
    for _ in range(NbIter):
        randomPortfolio = generateRandomPortfolio(Omega, k)
        randomAverage = getMeanChanges(randomPortfolio)
        if randomAverage > averageReturns:
            nbHigher += 1
    perc = round(nbHigher/NbIter*100)
    print(f"{k} asset(s) in {NbIter} random portfolio(s)")
    print("Average returns :", round(averageReturns))
    print(f"Percentage of random portfolios above the average : {perc}%")
    return perc
\end{lstlisting}

\begin{lstlisting}[language=Python]
# This function returns whether a random portfolio outperforms an average portfolio, 
# provided that 51% or more random portfolios outperform the average
def isRandomBetter(list, NbIter, k):
    perc = getPercentageHigherThanAverage(list, NbIter, k)
    if perc < 51:
        return False
    else:
        return True
\end{lstlisting}

\begin{lstlisting}[language=Python]
# Tests
print("Test 1")
print(isRandomBetter(Omega, 10, 10))
print("Test 2")
print(isRandomBetter(Omega, 10, 20))
print("Test 3")
print(isRandomBetter(Omega, 20, 10))
print("Test 4")
print(isRandomBetter(Omega, 20, 20))
print("Test 5")
print(isRandomBetter(Omega, 20, 30))
print("Test 6")
print(isRandomBetter(Omega, 30, 20))
print("Test 7")
print(isRandomBetter(Omega, 30, 30))
print("Test 8")
print(isRandomBetter(Omega, 30, 10))
print("Test 9")
print(isRandomBetter(Omega, 10, 30))
print("Test 10")
print(isRandomBetter(Omega, 40, 5))
\end{lstlisting}

\section{\texttt{isSMABetter($\Omega, n, r$)}}
\label{appendix:isSMABetter}

\begin{lstlisting}[language=Python]
# This function returns True if the average return of the SMA strategy
# is higher than the average of both hold and random strategies
def isSMABetter(Omega, n, r):
    validAssets = 0
    SMARets = []
    HoldRets = []
    RandomRets = []
    nbBetter = 0
    for asset in Omega:
        sma_return = getSMAReturn(asset, n, r)
        if not sma_return:
            continue
        else:
            SMARets.append(sma_return)
        hold_return = getHoldReturn(asset)
        if not hold_return:
            continue
        else:
            HoldRets.append(hold_return)
        random_return = getRandomReturn(asset)
        if not random_return:
            continue
        else:
            RandomRets.append(random_return)
        if sma_return > hold_return and sma_return > random_return:
            nbBetter += 1
        validAssets += 1
    
    sma_average = round(sum(SMARets) / len(SMARets))
    hold_average = round(sum(HoldRets) / len(HoldRets))
    random_average = round(sum(RandomRets) / len(RandomRets))
    print("Number of valid assets : ", validAssets)
    print("SMA average : ", sma_average)
    print("Hold average : ", hold_average)
    print("Random average : ", random_average)
    perc = round(nbBetter/validAssets*100)
    print(f"{perc}% of assets do better with SMA.")
    if perc < 50:
        return False
    else:
        return True
\end{lstlisting}

\section{\texttt{getHoldReturn(asset)}}
\begin{lstlisting}[language=Python]
# This function returns the return of the asset "asset" with the hold strategy
def getHoldReturn(asset):
    df = yf.download(asset, period = '2y', interval = '1d', progress=False)
    if df.empty:
        return False
    oneYear = df.loc['2021-01-01':'2022-01-01']
    s = pd.Series(list(oneYear['Close']))
    if oneYear.empty or len(oneYear) < 360 or not s[s.isin([0])].empty:
        return False
    else:
        start = oneYear.iloc[0]['Close']
        if start == 0:
            return False
        else:
            final = oneYear.iloc[-1]['Close']
            return round(((final-start)/start)*100)
\end{lstlisting}

\section{\texttt{getSMAReturn(asset, n, r)}}
\begin{lstlisting}[language=Python]
# This function returns the sum of daily returns
# of the asset "asset" with the SMA trading strategy
def getSMAReturn(asset, n, r):
    range = 1+(r/100)
    df = yf.download(asset, period = '2y', interval = '1d', progress=False)
    if df.empty:
        return False
    oneYear = df.loc['2021-01-01':'2022-01-01']
    s = pd.Series(list(oneYear['Close']))
    if oneYear.empty or len(oneYear) < 360 or not s[s.isin([0])].empty:
        return False
    else:
        oneYear['SMA'] = oneYear['Close'].shift(1).rolling(window=n).mean()
        oneYear['SMAhigh'] = oneYear['SMA']*range
        oneYear['SMAlow'] = oneYear['SMA']/range
        oneYear['Signal'] = 0
        oneYear.loc[oneYear['Close'] > oneYear['SMAhigh'], 'Signal'] = -1
        oneYear.loc[oneYear['Close'] < oneYear['SMAlow'], 'Signal'] = 1
        oneYear['Change'] = ((oneYear['Close']-oneYear['Close'].shift(1))/oneYear['Close'].shift(1))*100
        oneYear['DayReturn'] = oneYear['Change']*oneYear['Signal']
        ret = round(oneYear['DayReturn'].sum())
        return ret
\end{lstlisting}

\section{\texttt{getRandomReturn(asset)}}
\begin{lstlisting}[language=Python]
# This function returns the sum of daily returns
# of the asset "asset" with a random trading strategy
def getRandomReturn(asset):
    df = yf.download(asset, period = '2y', interval = '1d', progress=False)
    if df.empty:
        return False
    oneYear = df.loc['2021-01-01':'2022-01-01']
    s = pd.Series(list(oneYear['Close']))
    if oneYear.empty or len(oneYear) < 360 or not s[s.isin([0])].empty:
        return False
    else:
        oneYear['Signal'] = 0
        oneYear['Random'] = [random.randint(1,9) for _ in oneYear.index]
        oneYear.loc[oneYear['Random'] > 6, 'Signal'] = 1
        oneYear.loc[oneYear['Random'] < 4, 'Signal'] = -1
        oneYear['Change'] = ((oneYear['Close']-oneYear['Close'].shift(1))/oneYear['Close'].shift(1))*100
        oneYear['DayReturn'] = oneYear['Change']*oneYear['Signal']
        return round(oneYear['DayReturn'].sum())
\end{lstlisting}

\section{\texttt{getRandomPerc($\Omega$)}}
\label{appendix:getRandomPerc}
\begin{lstlisting}[language=Python]
# This function returns the percentage of assets that follow a random walk
def getPercRandom(Omega):
    nbRandom = 0
    nbTotal = 0
    for asset in Omega:
        time.sleep(1)
        df = yf.download(asset, period = 'max', interval = '1d', progress=False)
        if df.empty:
            continue
        s = pd.Series(list(df['Close']))
        if not s[s.isin([0])].empty or len(df) < 100:
            continue
        else:
            nbTotal += 1
            pval = adfuller(df['Close'])[1]
            if pval > 0.05:
                nbRandom +=1
    perc = nbRandom/nbTotal*100
    return perc
\end{lstlisting}

\section{\texttt{getAverageAccuracy($\Omega, n$)}}
\label{appendix:getAverageAccuracy}
\begin{lstlisting}[language=Python]
# This function returns the average accuracy percentage of our machine learning model
def getAverageAccuracy(Omega, n):
    accuracies = []
    for asset in Omega:
        df = yf.download(asset, period = '1y', interval = '1d', progress=False)
        df = df.drop(df[df['Volume'] == 0].index)
        df['RSI'] = ta.RSI(np.array(df['Close'].shift(1)), timeperiod=n)
        df['SMA'] = df['Close'].shift(1).rolling(window=n).mean()
        df['Corr'] = df['Close'].shift(1).rolling(window=n).corr(df['SMA'].shift(1))
        df['SAR'] = ta.SAR(np.array(df['High'].shift(1)), np.array(df['Low'].shift(1)), 0.2, 0.2)
        df['ADX'] = ta.ADX(np.array(df['High'].shift(1)), np.array(df['Low'].shift(1)), np.array(df['Close'].shift(1)), timeperiod=n)
        df['ATR'] = ta.ATR(np.array(df['High'].shift(1)), np.array(df['Low'].shift(1)), np.array(df['Close'].shift(1)), timeperiod=n)
        df['PH'] = df['High'].shift(1)
        df['PL'] = df['Low'].shift(1)
        df['PC'] = df['Close'].shift(1)
        df['O-O'] = df['Open'] - df['Open'].shift(1)
        df['O-C'] = df['Open'] - df['PC'].shift(1)
        df['Ret'] = (df['Open'].shift(-1) - df['Open']) / df['Open']
        for i in range(1, n):
            df['r%i' % i] = df['Ret'].shift(i)
        df.loc[df['Corr'] < -1, 'Corr'] = -1
        df.loc[df['Corr'] > 1, 'Corr'] = 1
        df = df.dropna()
        t = 0.8
        split = int(t*len(df))
        df['Signal'] = 0
        df.loc[df['Ret'] > df['Ret'][:split].quantile(q=0.66), 'Signal'] = 1
        df.loc[df['Ret'] < df['Ret'][:split].quantile(q=0.34), 'Signal'] = -1
        X = df.drop(['Close','Adj Close','Signal','High','Low','Volume','Ret'], axis=1)
        y = df['Signal']
        c = [10,100,1000,10000,100000,100000]
        g = [1e-4,1e-3,1e-2,1e-1,1e0]
        p = {'svc__C': c, 'svc__gamma': g, 'svc__kernel': ['rbf']}
        s = [('s', StandardScaler()), ('svc', SVC())]
        pp = Pipeline(s)
        rcv = RandomizedSearchCV(pp, p, cv = TimeSeriesSplit(n_splits=2))
        rcv.fit(X.iloc[:split], y.iloc[:split])
        c = rcv.best_params_['svc__C']
        k = rcv.best_params_['svc__kernel']
        g = rcv.best_params_['svc__gamma']
        cls = SVC(C = c, kernel = k, gamma = g)
        S = StandardScaler()
        cls.fit(S.fit_transform(X.iloc[:split]), y.iloc[:split])
        y_predict = cls.predict(S.transform(X.iloc[split:]))
        df['Pred_Signal'] = 0
        df.iloc[:split, df.columns.get_loc('Pred_Signal')] = pd.Series(
            cls.predict(S.transform(X.iloc[:split])).tolist())
        df.iloc[split:, df.columns.get_loc('Pred_Signal')] = y_predict
        df['Ret1'] = df['Ret'] * df['Pred_Signal']
        cr = classification_report(y[split:], y_predict, output_dict=True)
        accuracies.append(cr['accuracy'])
    return round(sum(accuracies) / len(accuracies) * 100)
\end{lstlisting}

\section{NLP Trading Bot}
\label{appendix:nlp}
\begin{lstlisting}[language=Python]
import tweepy
import time
from textblob import TextBlob
import yfinance as yf

# Authentication
key = "" 
csecret = "" 
atoken = "" 
atsecret = ""
nb = 500 
keywords = ["BTC", "#BTC", "Bitcoin"]

auth = tweepy.OAuthHandler(ckey, csecret)
auth.set_access_token(atoken, atsecret)
api2 = tweepy.API(auth, wait_on_rate_limit=True, wait_on_rate_limit_notify=True)

def perc(a, b):
    temp = 100 * float(a) / float(b)
    return format(temp, '.2f')

def get_current_price(symbol):
    ticker = yf.Ticker(symbol)
    todays_data = ticker.history(period='1d')
    return todays_data['Close'][0]

def get_twitter_BTC():
    ratios = 0
    for keyword in keywords:
        tweets = tweepy.Cursor(api2.search, q=keyword, lang="en").items(nb)
        pos = 0
        neg = 0
        for tweet in tweets:
            analysis = TextBlob(tweet.text)
            if 0 <= analysis.sentiment.polarity <= 1:
                pos += 1
            elif -1 <= analysis.sentiment.polarity < 0:
                neg += 1
        pos = perc(pos, nb)
        neg = perc(neg, nb)
        if float(neg) > 0:
            ratio = float(pos) / float(neg)
        else:
            ratio = float(pos)
        ratios += ratio
    return ratios

if __name__ == "__main__":
    for k in range(1000):
        score = get_twitter_BTC()
        min1 = score + (score * 30 / 100)
        time.sleep(60*5)
        new_score = get_twitter_BTC()
        if new_score > min1:
            btc_price = get_current_price("BTC-USD")
            buy = "\nBUY : " + str(btc_price)
            with open("output.txt", "a") as f:
                f.write(buy)
            time.sleep(60*5)
            new_new_score = get_twitter_BTC()
            min2 = new_score - (new_score * 30 / 100)
            if new_new_score < min2:
                new_btc_price = get_current_price("BTC-USD")
                sell_at = " SELL : " + str(new_btc_price)
                trade_profit = new_btc_price - btc_price
                perc_profit = trade_profit / btc_price * 100
                perc_profit_round = round(perc_profit, 3)
                sell_message = sell_at + " | " + " Profit = " + str(perc_profit_round) + " %"
                with open("output.txt", "a") as f:
                    f.write(sell_message)
                time.sleep(60*5)
        else:
            time.sleep(60*5)
\end{lstlisting}

\pagebreak
\bibliographystyle{apalike}
\bibliography{bib.bib}
%%%%%%%%%%%%%%%%%%%%%%%%%%%%%%%%%%%%%%%%%%%%%%%%%%%%%

\end{document}